\documentclass[prl,twocolumn,amsmath,amssymb,superscriptaddress,unsortedaddress]{revtex4}

\usepackage{epsf}
\usepackage{graphicx}
\usepackage[usenames]{color}

\begin{document}

\title{Quadratic Fermi Node in a 3D Strongly Correlated Semimetal}

\author{Takeshi Kondo}
\affiliation{ISSP, University of Tokyo, Kashiwa, Chiba 277-8581, Japan}

\author{M.~Nakayama} 
\affiliation{ISSP, University of Tokyo, Kashiwa, Chiba 277-8581, Japan}

\author{R.~Chen} 
\affiliation{Physics Department, University of California, Santa Barbara, California 93106, USA}
\affiliation{Physics Department, University of California, Berkeley, California 94720, USA}
\affiliation{Molecular Foundry, Lawrence Berkeley National Laboratory, Berkeley, California 94720, USA}

\author{J.J.~Ishikawa} 
\affiliation{ISSP, University of Tokyo, Kashiwa, Chiba 277-8581, Japan}

\author{E.-G.~Moon} 
\affiliation{Physics Department, University of California, Santa Barbara, California 93106, USA}
\affiliation{Department of Physics, Korea Advanced Institute of Science and Technology, Daejeon 305-701, Korea}

\author{T.~Yamamoto} 
\affiliation{ISSP, University of Tokyo, Kashiwa, Chiba 277-8581, Japan}

\author{Y.~Ota} 
\affiliation{ISSP, University of Tokyo, Kashiwa, Chiba 277-8581, Japan}

\author{W.~Malaeb} 
\affiliation{ISSP, University of Tokyo, Kashiwa, Chiba 277-8581, Japan}
\affiliation{Physics Department, Faculty of Science, Beirut Arab University, Beirut, Lebanon}

\author{H.~Kanai} 
\affiliation{ISSP, University of Tokyo, Kashiwa, Chiba 277-8581, Japan}

\author{Y.~Nakashima} 
\affiliation{ISSP, University of Tokyo, Kashiwa, Chiba 277-8581, Japan}

\author{Y.~Ishida} 
\affiliation{ISSP, University of Tokyo, Kashiwa, Chiba 277-8581, Japan}

\author{R.~Yoshida} 
\affiliation{ISSP, University of Tokyo, Kashiwa, Chiba 277-8581, Japan}

\author{H.~Yamamoto} 
\affiliation{ISSP, University of Tokyo, Kashiwa, Chiba 277-8581, Japan}

\author{M.~Matsunami} 
\affiliation{UVSOR Facility, Institute for Molecular Science, Okazaki 444-8585, Japan}
\affiliation{Energy Materials Laboratory, Toyota Technological Institute, Nagoya 468-8511, Japan}

\author{S.~Kimura} 
\affiliation{UVSOR Facility, Institute for Molecular Science, Okazaki 444-8585, Japan}
\affiliation{Graduate School of Frontier Biosciences, Osaka University, Suita, Osaka, 565-0871, Japan }

\author{N.~Inami} 
\affiliation{Institute of Materials Structure Science, High Energy Accelerator Research Organization (KEK),  Tsukuba, Ibaraki 305-0801, Japan}

\author{K.~Ono} 
\affiliation{Institute of Materials Structure Science, High Energy Accelerator Research Organization (KEK),  Tsukuba, Ibaraki 305-0801, Japan}

\author{H.~Kumigashira} 
\affiliation{Institute of Materials Structure Science, High Energy Accelerator Research Organization (KEK),  Tsukuba, Ibaraki 305-0801, Japan}

\author{S.~Nakatsuji} 
\affiliation{ISSP, University of Tokyo, Kashiwa, Chiba 277-8581, Japan}
\affiliation{PRESTO, Japan Science and Technology Agency (JST), 4-1-8 Honcho Kawaguchi, Saitama 332-0012, Japan}

\author{L.~Balents} 
\affiliation{Kavli Institute for Theoretical Physics, Santa Barbara, California 93106, USA}

\author{S.~Shin} 
\affiliation{ISSP, University of Tokyo, Kashiwa, Chiba 277-8581, Japan}

\date{\today}

\maketitle

{\bf Strong spin-orbit coupling fosters exotic electronic states such as topological insulators and superconductors, but the combination of strong spin-orbit and strong electron-electron interactions is just beginning to be understood.  Central to this emerging area are the $5d$ transition metal iridium oxides.  Here, in the pyrochlore iridate Pr$_2$Ir$_2$O$_7$, we identify a nontrivial state with a single point Fermi node protected by cubic and time-reversal symmetries, using a combination of angle-resolved photoemission spectroscopy and first principles calculations.  Owing to its quadratic dispersion, the unique coincidence of four degenerate states at the Fermi energy, and strong Coulomb interactions, non-Fermi liquid behavior is predicted, for which we observe some evidence.  Our discovery implies that Pr$_2$Ir$_2$O$_7$ is a parent state that can be manipulated to produce other strongly correlated topological phases, such as topological Mott insulator, Weyl semi-metal, and quantum spin and anomalous Hall states.
}

Following the discovery of topological insulators \cite{Hasan_TI,Fu_TI,Moore_TI,Ando}, the next frontier is the regime in which both spin-orbit coupling and correlation effects are strong \cite{Kim_PRL,Nakatsuji_Nature,YongBaek_PRB,Leon_NP, witczak2013correlated, Leon_PRL2013,Eu227_weyl}.
Theory has suggested that the pyrochlore iridates, a family of cubic  $5d$ transition metal oxides \cite{maeno,Matsuhira_Ln2Ir2O7}, may realize both band inversion, the essential ingredient of topological insulators, and strong correlations \cite{Eu227_weyl,Leon_NP}. Empirical evidence for the latter is plentiful.  Notably, Pr$_2$Ir$_2$O$_7$ appears to be proximate to an interaction-driven antiferromagnetic quantum critical point  tuned by $A$ site ionic radius, which is located between the two ions with largest radii, $A$ = Pr and $A$ = Nd \cite{Nakatsuji_PRL,Matsuhira_Ln2Ir2O7}. 
It is the only compound among the series in which the iridium electrons remain paramagnetic and itinerant down to the lowest measured temperatures.  
It displays ``bad metallic'' behavior and nontrivial spontaneous Hall transport, suggesting strong correlations \cite{Nakatsuji_PRL,machida2007,Nakatsuji_Nature}. Moreover, recent thermodynamic measurements have revealed zero-field quantum criticality without tuning \cite{tokiwa2013}. 

The phenomenological suggestion of Ref.\cite{Leon_PRL2013}, whose implications are summarized in Fig. 1, is that the Fermi surface of Pr$_2$Ir$_2$O$_7$ contains a single Fermi node at the $\Gamma$ point, which emerges as the touching point of two quadratically dispersing ``conduction'' and ``valence'' bands \cite{QuadraticBandTouching}.  The presence of this touching is actually required by symmetry and group theory (the quadruplet at the zone center lies in the $\Gamma_8$ representation of the double group of $O_h$), but its location directly at the Fermi energy was an ad-hoc theoretical assumption.  
If the assumption is correct, Pr$_2$Ir$_2$O$_7$ becomes a strongly correlated analog of HgTe \cite{HgTe_science,HgTe_ARPES}, which has a mathematically identical quadratic node at the Fermi energy, and implies that Pr$_2$Ir$_2$O$_7$ should be tunable into various topological states (see Fig. 1).  
Furthermore, theory has predicted that the quadratic nodal semimetal itself is fundamentally altered by long-range Coulomb interactions (negligible in HgTe due to the large dielectric constant, but not so here), becoming a non-Fermi liquid state.   Thus the Fermi node, if correct, means that Pr$_2$Ir$_2$O$_7$ is a natural parent material for strongly interacting topological phases and non-Fermi liquid states.  

In this paper, we focus on the phenomena of band inversion, and present theoretical calculations and experimental angle-resolved photoemission spectroscopy (ARPES) spectra which support it in the form of electronic structure with a node at the Fermi energy.  We also observe strongly temperature dependent single-particle spectral weight and lineshape structure in ARPES which suggest electronic correlations and coupling to collective modes.

In support of this proposition, we first present ab initio electronic structure calculations in the paramagnetic state.  As detailed in the Supplementary Fig. 1, we show that the quadratic band touching systematically approaches the Fermi level with increasing $A$ site ionic radius, reaching it for $A$ = Pr.  The corresponding band dispersion is shown in Fig. 2g along the high symmetry lines. The nodal Fermi point and quadratic band touching at $\Gamma$ is clearly visible at the Fermi energy. The band width is narrower in energy than that of HgTe \cite{HgTe_calc} by one order of magnitude, reflecting the localized nature of the $5d$ orbitals in  Pr$_2$Ir$_2$O$_7$.
Theoretical uncertainty remains, however, as discussed in the Supplementary Note 1 (also see Supplementary Figs. 1 and 2), so direct experimental evidence for the unique Fermi node state is strongly desired.

Angle-resolved photoemission spectroscopy is a powerful technique to directly observe the electronic structure of matter \cite{Kiss,kimura2010samrai}.  One incident photon energy corresponds to one $k_z$ value in solid, thus the momentum space observed at a fixed photon energy is limited to a $k_x$-$k_y$ plane at a fixed $k_z$.  To locate the $\Gamma$-point, therefore, sweeping the photon energy is required.  Figures 2a and 2b shows the photos of    
the cleavage surface with the (111) plane measured by ARPES and the high-quality single crystal we used, respectively.  
While the cleaved surface looks rough in the photo, the SEM image of it  (Figs. 2c and 2d) exhibits  
very flat parts in the multiple locations, which are large enough compared with the photon beam size ($\sim$150$\mu$m, marked with a yellow circle in Fig. 2d). 
Several $k_x$-$k_y$ sheets perpendicular to $k_{(111)}$ measured at different photon energies are colored in Fig. 2f.  The band dispersions obtained by the ab initio calculation for these sheets are plotted in Figs. 2h-2j. 
The experimentally obtained energy dispersion is expected to have a large gap 
when the $k_{(111)}$ location is far from the $\Gamma$.   As $k_{(111)}$ is reduced, the gap decreases and eventually vanishes at $\Gamma$, where the two parabolic dispersions touch at $E_{\rm F}$ (Fig. 2j).  

In order to observe the Fermi node, we measured the ARPES spectra at various photon energies: $h\nu = 7$ eV (a laser source) and 21.2 eV (a He lamp) in the lab system, and $h\nu = 8\sim 18$ eV and $39\sim 60$ eV from two different synchrotron facilities.  The results for the 1st Brillouin zone are shown in Fig. 3d, where the energy distribution curves (EDCs) along $k_x$ (defined in Fig. 3a) are plotted.  The observed momentum cut shifts with photon energy along the $k_{(111)}$ axis within a momentum sheet crossing $\Gamma$, L, and K points (colored dashed lines in Fig. 3a). The spectra are symmetrized to remove the effect of the Fermi cut-off: the EDC is flipped about $E_{\rm F}$ and added to the original one.  This method is widely accepted as a means of determining the presence of a gap (or a gap node).  Small but clear quasiparticle peaks are seen for all the spectra, which allows us to determine the energy dispersion, as marked by bars and dotted curves.  At $h\nu=7$ eV (see Fig. 3d), a large gap of $\sim 20$ meV is opened at $k_x=0$ (green curve), and a parabolic dispersion is obtained. The most significant finding is that the parabolic dispersion moves toward the $E_{\rm F}$ with increasing incident photon energy (or increasing $k_{(111)}$), and it eventually touches $E_{\rm F}$ at $h\nu=$10 eV.  This behavior is more clearly demonstrated in Figs. 3b and 3c, where the band dispersion determined from the peak energies of spectra at 7, 8, 9, and 10 eV are plotted.  As the photon energy is further increased, the dispersion moves away from $E_{\rm F}$ again, following the quadratic dispersion along $k_{(111)}$ as shown in Fig. 3e (also see Supplementary Figs. 3 and 4, and Supplementary Note 2).
 We have also examined the dispersion along a different momentum sheet crossing $\Gamma$, L and W points (see Supplementary Fig. 5 and Supplementary Note 2) for a different piece of sample. This time the photon energy was swept by a finer step ($\le$ 0.5 eV), and the Fermi node was again detected at $h\nu=$10.5 eV.  While it is not possible to eliminate a slight uncertainty in the exact photon energy which yields $k_{(111)}=0$, owing to the broad shape of the spectral peak, our data show that the 3D band structure of Pr$_2$Ir$_2$O$_7$ has the theoretically predicted Fermi point at the momentum reached by $h\nu \sim10$ eV, which is thus assigned to be $\Gamma$.  Other scans of different $k_{(111)}$ values up to the L point in the 1st Brillouin zone, which is reached at $h\nu=18$ eV, revealed no other states touching or crossing $E_{\rm F}$ (see Supplementary Fig. 3).  This absence of other bands crossing $E_{\rm F}$ is another consistency condition on the Fermi node model, which requires this situation by state counting and charge neutrality.

Here we validate our conclusion by making several further checks.  First, we investigate another Brillouin zone to verify the required repetition of the Fermi nodal state along $k_{(111)}$. We used higher photon energies, corresponding to the 3rd Brillouin zone, and found the Fermi node at the expected $\Gamma$ point ($h\nu = 52$ eV) (see Supplementary Fig. 6).
In the $k_x-k_y$ sheet at $h\nu = 52$ eV (Fig. 4a and 4b), the ARPES intensities at $E_{\rm F}$ becomes strongest at the zone center, and the band touching at $E_{\rm F}$ is confirmed in the dispersion maps along $k_x$ and $k_y$ (Fig. 3d) and the corresponding symmetrized EDCs (Fig. 4f, also see the raw EDCs in Supplementary Fig. 7). In contrast, these features are missing in the $k_x-k_y$ sheet across L point ($h\nu = 39$ eV) (see Fig. 4c), where a rather flat, gapped dispersion is observed (Figs. 4e, and 4g).

Second, we demonstrate that our conclusion is insensitive to the different analytic schemes. 
This is significant especially because the symmetrization technique is relevant for the particle-hole symmetric state, which is unknown in Pr$_2$Ir$_2$O$_7$. 
 Accordingly we have tested another widely-used measure of dividing the ARPES spectra by the Fermi function. 
 Figure 5c plots the EDCs along a momentum cut crossing $\Gamma$ (light blue arrow in the inset of Fig. 5d) measured at various temperatures.  Instead of symmetrization, the curves are divided by the energy-resolution-convoluted Fermi function at the measurement temperatures to remove the effect of the Fermi cut-off.  It is clearly seen that the quadratic dispersion touches $E_{\rm F}$, in agreement with the earlier analysis. In  Fig. 5e,  
 we compare the dispersions determined by the symmetrization and the FD-devision methods, and confirm the consistency between the two (see Supplementary Fig. 8 and Supplementary Note 3 for more details). 

In Fig. 3b, we compare our experimental results near $\Gamma$ with the ab-initio dispersion (gray curve), which has a purely quadratic shape of the electronic structure given by $\varepsilon (k) \propto {k^2}$.  Close to $E_{\rm F}$, we find an  agreement between the two, demonstrating that the quadratic curve (light blue dotted line) fits well to our data with almost the same effective mass, $m_{\rm eff}=6.3m_0$ ($m_0$ is the mass of a free electron). 
On the other hand, for  energy below -0.012 eV, the measured dispersion deviates remarkably from the parabolic shape. This contrasts to the calculation, which matches with $\varepsilon (k) \propto {k^2}$ up to much higher binding energies.  
It is possible that the deviation is due to correlation effects beyond the band calculation. 
 Intriguingly, the total band width in the occupied side is estimated to be $\sim$40 meV (arrows in Figs. 4f and 4g), which is actually much narrower than that ($>100$ meV) of the band calculation (see Fig. 2g and Supplementary Figs. 9e and 9h). 
Band narrowing relative to density functional theory is indeed a well-known characteristic of correlated electrons. 
If that is the case, however, the agreement of the effective mass around $\Gamma$ between the data and calculation would be a coincidence.
This is understandable, considering that the band shape of  Pr$_2$Ir$_2$O$_7$ with comparable energy scales between the spin-orbit interaction and the electron correlations is sensitive to different calculation methods (see Supplementary Note 4, and Supplementary Figs. 1 and 9). 

Another possible cause of the discrepancy between the data and calculations
is the strong coupling of electrons to the bosonic modes,
which  also could significantly renormalize the band shape. 
Indeed, the peak-dip-hump shape, which is a characteristic signature of strong mode coupling, is seen in EDCs (Fig. 3e), being consistent with this scenario. One of candidates for the bosonic modes is the phonons, which are usually  coupled to the correlated  systems very strongly.  Another candidate is suggested by the similarity of the slightly distorted band shape to measurements in graphene, where it was attributed to electron-plasmon coupling \cite{Rotenberg_NP}.  Namely, the origin could be the same in the two cases: emission of collective modes through vertical interband transitions becomes possible above a threshold energy, modifying the spectrum.  

The strongly correlated feature of $5d$ electrons of Pr$_2$Ir$_2$O$_7$ should be observed in the ARPES spectra.  We find such a signature in the temperature evolution of the spectral shape.  Figures 5a and 5b plot Fermi-function divided EDCs at $\Gamma$ ($h\nu = 10.5$ eV) and off $\Gamma$ ($k_{(111)}=0.31$ $\rm{\AA}^{-1}$  reached at $h\nu =21.2$ eV).  The sharp peak clearly seen in the low temperature spectra is strongly suppressed at elevated temperatures. The  behavior is clearly more dramatic than the thermal broadening effects observed in the other strongly collated systems such as the well-studied cuprates, which have the ``marginal" Fermi liquid state \cite{Varma,Kondo_resistivity}. 
We note that the peak suppression in the present data is accelerated across $\sim 100$K, differently from  
the typical thermal broadening with a gradual increase over a wider temperature range. 
We find that, associated with the suppression, a large portion of spectral weight above $E \simeq-0.1$ eV is transferred to higher binding energies in Pr$_2$Ir$_2$O$_7$.  
A plausible origin for it is the polaronic effects, which could become crucial in the purely screened electronic systems, as also proposed for the other strongly correlated systems such as the perovskite iridates, manganites, and lightly doped cuprates \cite{Sr3Ir2O7,Dessau,Shen}.
These features are compatible with the non-Fermi liquid behavior predicted theoretically for the Fermi node phase in Ref.\cite{Leon_PRL2013,savary2014new}, and may also be related to recent observations of quantum critical behavior in thermodynamic measurements of Pr$_2$Ir$_2$O$_7$ \cite{tokiwa2013}. 
However,  a fuller identification of the non-Fermi liquid state and explication of its physics in Pr$_2$Ir$_2$O$_7$ requires higher resolution data and more elaborate analysis, beyond the scope of this paper.

In passing, we point out that broad spectral weight emerges beyond $E_{\rm F}$ as seen in the Fermi-function divided image for the 75K data (arrow in Fig. 5f).  The related spectral intensity is obtained off $\Gamma$, showing an upturn behavior beyond $E_{\rm F}$ (arrow in Fig. 5b and Supplementary Figs. 10 and 11, and see Supplementary Notes 5 and 6).  While the strong suppression of quasiparticle peaks at elevated temperatures prevents us from a definitive determination of the conduction band dispersion,
the observation of spectral weight above $E_{\rm F}$ is compatible with the predicted existence of a quadratic band touching on the unoccupied side (Fig.  5d).

Prior measurements \cite{machida2007,Nakatsuji_Nature,Balicas} showing ferromagnetic spin-ice type correlations amongst the Pr moments below 2 K may be explained due to unconventional ferromagnetic RKKY interactions arising from the point-like Fermi surface. The Fermi node also leads to strong sensitivity to small time reversal breaking perturbations, producing Weyl points close to the Fermi energy \cite{Eu227_weyl} and a gigantic anomalous Hall effect. This is also in accord with the experimental fact that Pr$_2$Ir$_2$O$_7$ was the first material found to exhibit a large spontaneous Hall effect in a spin liquid state at zero field \cite{Nakatsuji_Nature,Balicas}.

Our results suggest that tuning of a unique quantum critical point between an antiferromagnetic Weyl semimetal and the non-Fermi liquid nodal phase may be possible by alloying or hydrostatic pressure. Correlated topological phases and device applications with iridate films could be accessed by controlling the strain-induced breaking of the cubic crystal symmetry and size quantization (subband formation) in quantum well structures.  We indeed verified theoretically the opening of a significant topological gap with uniaxial compression along the $\langle 111\rangle$ axis by first principles calculations (see Supplementary Fig. 12 and Supplementary Note 7).  This analysis moreover shows the presence of three two-dimensional surface Dirac cones in this TI state  (see Supplementary Fig. 13, Supplementary Table 1, and Supplementary Note 7).  It will be exciting to investigate whether correlations, neglected in density function theory, lead to spontaneous time reversal breaking at surfaces \cite{PhysRevB.85.121105,PhysRevB.81.020411} with fractional excitations \cite{qi2008fractional}, or surface topological order \cite{PhysRevB.88.115137,metlitski2013symmetry,chen2013symmetry,bonderson2013time}, both of which have been predicted theoretcally.  

\textbf{Methods} 

\textbf{Samples and ARPES experiments.} 
Single crystals of Pr$_2$Ir$_2$O$_7$ with 1 mm$^3$ in size were grown using a flux method.
The sample surface with the (111) plane was prepared by cleaving the single crystal {\it in situ} with a top post clued on the crystal. 

The ARPES experiments were performed at 
 BL7U of Ultraviolet Synchrotron Orbital Radiation  (UVSOR) facility  ($h\nu = 8\sim 18$ eV) with a MBS A-1 electron analyzer \cite{kimura2010samrai}, BL28A of Photon Factory, KEK, ($h\nu = 39\sim 60$ eV) with a Scienta SES2002 electron analyzer, 
 and in our laboratory using a system consisting of a Scienta R4000 electron analyzer equipped with 
 a 6.994 eV laser (the 6th harmonic of Nd:YVO$_4$ quasi-continuous wave with a repetition rate of 240 MHz) \cite{Kiss} and He discharge lamp (HeI$\alpha$, $h\nu$ = 21.2 eV). 
The overall energy resolution in the ARPES experiment was set to $\sim2$meV and $\sim6$ meV in the lab system with a laser and He lamp, respectively,  and 
$\sim 15$ meV for the synchrotron facility data.   
The sample orientation was determined with a Laue picture taken before the ARPES experiment.

The intrinsic $k_z$ broadening, $\delta {k_z}$, is inversely proportional to the photoelectron escape depth  $\lambda$ ($\delta  k_z =1/ \lambda $). We used low photon energies to find the nodal point in the 1st Brillouin zone, which enables the bulk sensitive measurements. For example, the photon energy corresponding to the $\Gamma$ point is around 10 eV, at which the $\lambda$ value is estimated to be $\sim$20 $\rm{\AA}$ according to the ``universal curve".   It translates into $\delta {k_z}$ with $\sim$5 $\%$ of the BZ size, which is rather small and sufficient to validate our results. In the  Supplementary Fig. 4,  we demonstrate that the $\delta {k_z}$ (or $\delta {k_{(111)}}$) is small enough to resolve the $E$ vs $k_{\rm {(111)}}$ relation with a 1-eV step of photon energy. 
We also used higher photon energies of $\sim$50eV to investigate the 3rd Brillouin zone. In this case, the $\lambda$ value becomes small ($\sim$5 $\rm{\AA}$), resulting in a relatively large $\delta {k_z}$ ($\sim$20 $\%$ of the BZ). Nevertheless, we note that the energy dispersion in Pr$_2$Ir$_2$O$_7$ is very weak especially around the $\Gamma$ point (quadratic Fermi node), thus the effects of $k_z$ broadening on the quasiparticle peaks should not be critical in our study. 

\textbf{Band calculations.} 
Calculations were carried out using both the standard generalized gradient approximation (GGA)  and the Tran-Blaha modified Becke-Johnson (TB-mBJ) \citep{TB-mBJ} exchange potential as implemented in Wien2k, for a series of different A site ions. An RKmax parameter 7.0 was chosen and the wave functions were expanded in spherical harmonics up to $l_{\text max}^{\text wf}=10$ inside the atomic spheres and  $l_{\text max}^{\text pot}=4$ for non-muffin tins. Bulk A$_2$Ir$_2$O$_7$ (A=rare earth element) has a cubic crystal structure with the space group Fd$\bar 3$m. The experimental lattice parameters were used for A= Pr, Nd, Eu and Y in both GGA and TB-mBJ paramagnetic calculations, and spin-orbit coupling was applied to both the heavy rare earth element and the Ir electrons. The paramagnetic GGA and TB-mBJ calculations put the 4f states of the rare earth element at the Fermi energy; to avoid this, since the 4f electrons are highly localized, their potential is shifted by a constant.

The TB-mBJ method is believed to produce improved results for small band gap systems \cite{meritlimit-mBJ}, and has been widely used in studies of topological insulators \cite{feng2010halfTI}.  Both methods (GGA and TB-mBJ) yield almost indistinguishable results away from the Fermi energy, with a slight decrease of band-width in the TB-mBJ calculation, and small differences near $E_F$, as shown in Supplementary Fig. 1.  We observed the symmetry-required nodal band touching at $\Gamma$ for all calculations, but the Fermi energy is shifted from the nodal point by an amount that decreases with increasing A (A= Y, Eu, Nd and Pr) site ionic radius in both methods \cite{Onoda}. This occurs because of accidental crossing of states near the L point, where the valence band rises and approaches the Fermi energy, especially in the smaller rare earths.  In general, the GGA calculation underestimates the effects of correlations, which tend to narrow the bands and to push occupied states deeper below the Fermi energy.  Hence our expectation is that the accidental crossing near L, which is responsible for the Fermi level shift in GGA, will be suppressed by correlations.  The TB-mBJ method may be regarded as a crude way to do this.  Indeed, in the presumed more accurate TB-mBJ method, all the paramagnetic band structures for the A$_2$Ir$_2$O$_7$ series show smaller shifts of the Fermi level compared to their GGA counterparts. In the case of A=Pr, we find the shift vanishes and the nodal point occurs precisely at the Fermi energy. While there is a universal trend of smaller shift of the Fermi energy as rare earth ionic radius increases for the rare earth elements we have tested, the small differences between GGA and TB-mBJ suggest that some theoretical uncertainty remains.  We note, however, that we expect methods that include correlations more accurately, such as LDA+DMFT, are likely to further suppress the states near the Fermi energy at L even beyond TB-mBJ, favoring placing the Fermi level precisely at the node.



 {\bf Acknowledgements} \\
 This work was supported by JSPS KAKENHI (Nos. 24740218, 25220707, and 25707030), by the Photon and Quantum Basic Research
Coordinated Development Program from MEXT,  by PRESTO, Japan Science and Technology Agency,  Grants-in-Aid for Scientific Research (No. 25707030), by Grants-in-Aids
for Scientific Research on Innovative Areas (15H05882, 15H05883), and Program for Advancing Strategic International Networks to Accelerate the Circulation of Talented Researchers (No. R2604) from the Japanese Society for the Promotion of Science.
L.B. and R.C. were supported by DOE grant DE-FG02-08ER46524, and E.-G. M. was supported by the MRSEC Program of the NSF under Award No. DMR 1121053. We thank D. Hamane for technical assistance in the SEM measurement. 
The use of the facilities (the Materials Design and Characterization Laboratory and the Electronic Microscope Section) at the Institute for Solid State Physics, The University of Tokyo, is gratefully acknowledged. 



\begin{figure*} \includegraphics[width=5.5in]{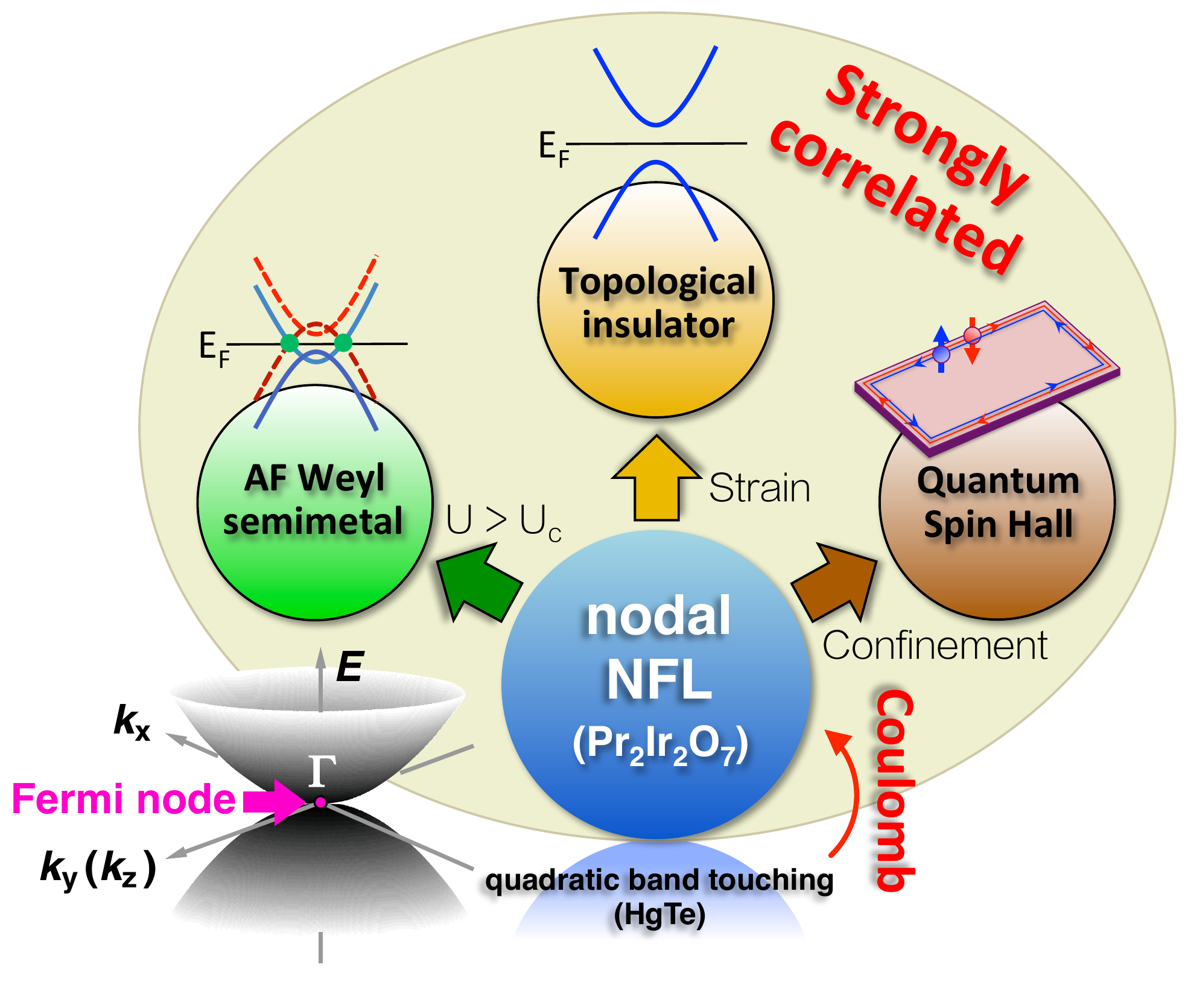} 
\caption{{\bf Quadratic Fermi node state of Pr$_2$Ir$_2$O$_7$ tunable into interacting topological phases.} In the lower part of the diagram, the bottom half of the blue circle and its reflection form a caricature of the quadratically dispersion conduction and valence bands touching at the zone center, while at the same time the darker blue upper circle suggests how Pr$_2$Ir$_2$O$_7$, with non-negligible Coulomb interactions, is a strongly correlated non-Fermi liquid analog of HgTe, shown as a pale blue reflection.  Arrows indicate the perturbations which convert the nodal non-Fermi liquid state to diverse topological phases: an antiferromagnetic Weyl semimetal should be produced in bulk materials by alloying or hydrostatic pressure,  uniaxial strain yields a three-dimensional topological insulator, and two-dimensional confinement produces a quantum spin Hall state. The outer circle reminds us that all the states produced in these ways retain strong correlations, and hence are excellent candidates for observing non-trivial surface phases different from those of band theory, as discussed in the text.} 
\label{fig1}
\end{figure*}

\clearpage

\begin{figure*} \includegraphics[width=5.3in]{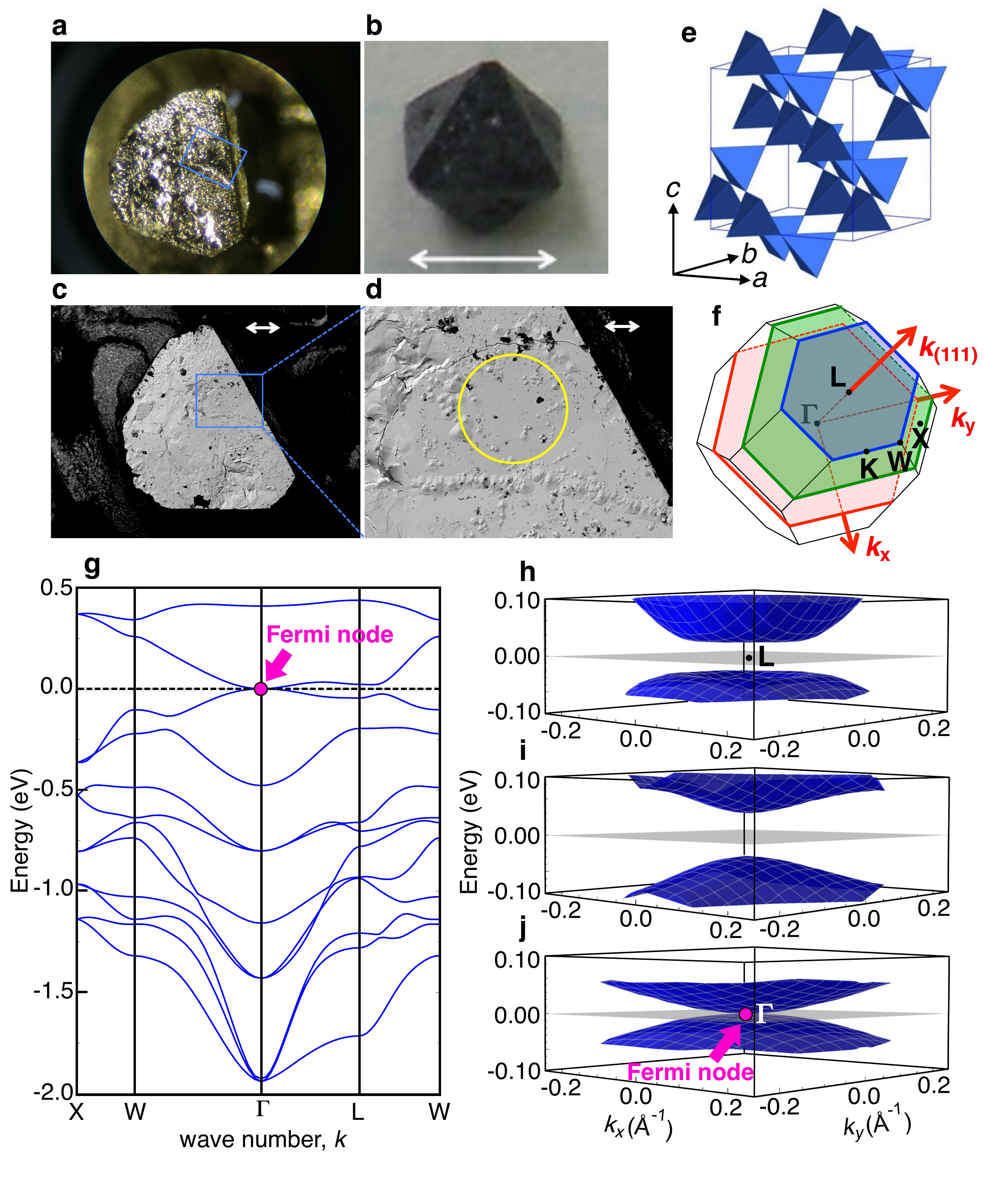} 
\caption{ {\bf Single crystal and first principles band calculation for Pr$_2$Ir$_2$O$_7$.} 
{\bf a,b,}  Photos of the cleaved (111) surface measured by ARPES and single crystal we used, respectively.
The top surface with a triangle shape is the (111) plane. 
{\bf c,} The SEM image for the cleaved surface of {\bf a}.  {\bf d,} The same image as {\bf c},
but magnified in a small region marked by a blue square in {\bf a} and {\bf c}.
The  arrows in {\bf b}, {\bf c}, and {\bf d} indicate dimensions of 1 mm, 200 $\mu$m, and 50 $\mu$m, respectively.
{\bf e,} Crystal structure of Pr$_2$Ir$_2$O$_7$ showing the pyrochlore lattice of Ir (or Pr) atoms.
{\bf f,} Brillouin zone.  {\bf g,} Band dispersion along high symmetry lines obtained by the first principles calculation.  Band dispersions in three $k_x-k_y$ planes (colored in {\bf f}) perpendicular to the $k_{(111)}$ direction: crossing L ({\bf h}), crossing $\Gamma$ ({\bf j}), and crossing between these two points ({\bf i}). The Fermi node is indicated with a magenta circle and arrow in  {\bf g} and  {\bf j}. }
\label{fig1}
\end{figure*}

\begin{figure*}
\includegraphics[width=5.5in]{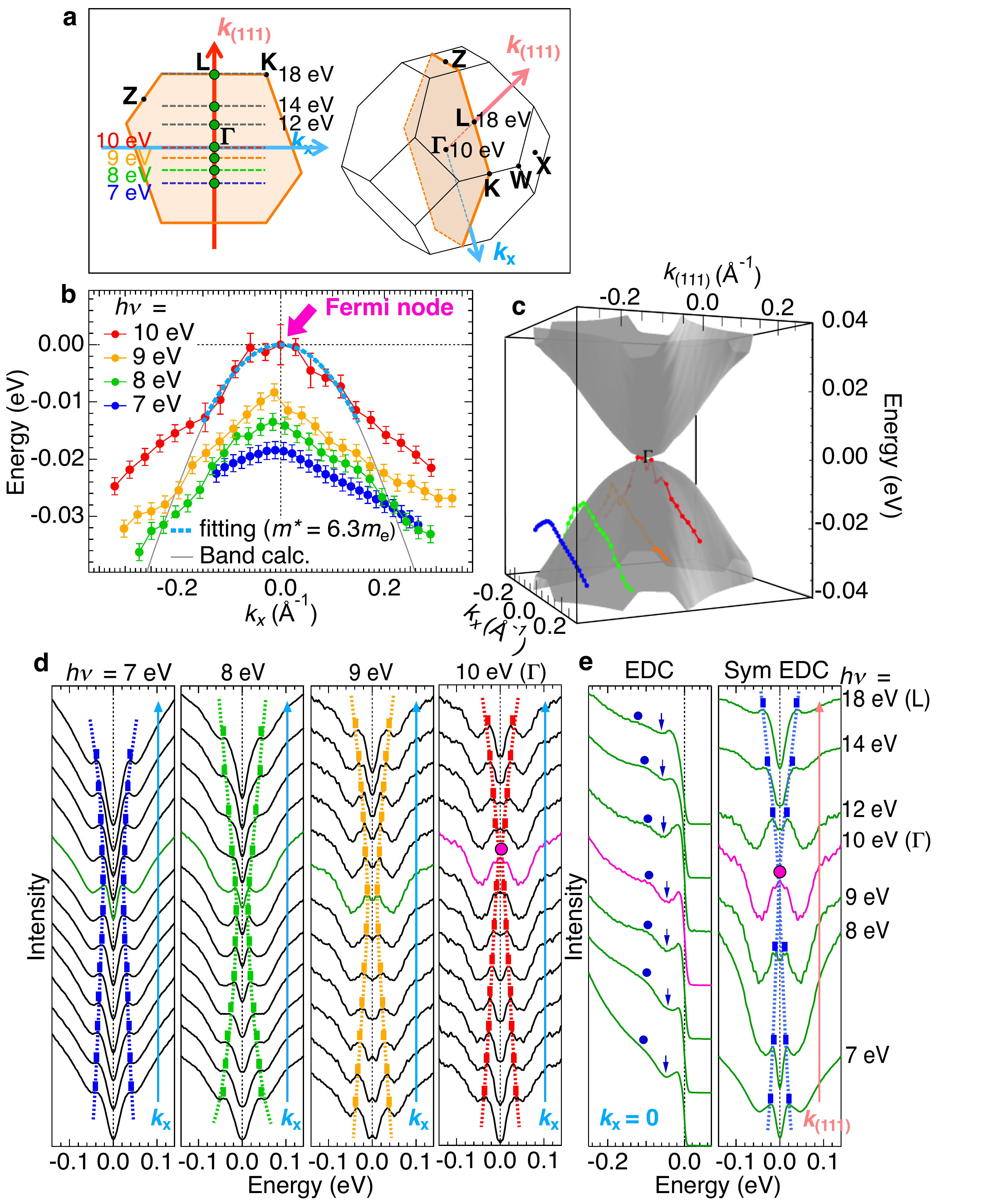}
\caption{{\bf ARPES spectra revealing a quadratic Fermi node in the 3D band of Pr$_2$Ir$_2$O$_7$.}
{\bf a,} Brillouin zone, showing  a momentum sheet along which  ARPES data were measured. The momentum cuts at $h\nu=$ 10 eV  and 18 eV crosses  the $\Gamma$ and L points in the 1st Brillouin zone.
{\bf b,} Energy dispersions along the $k_x$ direction measured at $h\nu=$ 7, 8, 9, and 10 eV. 
The corresponding momentum cuts are indicated in {\bf a} by dashed colored lines. 
The band dispersion obtained by the first principles band calculation is superimposed (gray curve). 
The data close to $E_{\rm F}$ is fitted by a parabolic function, $\varepsilon (k) \propto {k^2}$ (light blue dotted curve). 
The estimated effective mass at $\Gamma$, $m_{\rm eff}=6.3m_0$ ($m_0$:  free electron mass), is in agreement with the band calculation. 
{\bf c,} The calculated band dispersion in the $k_x - k_{(111)}$ sheet, painted with orange  in {\bf a}.  On it, the ARPES data in  {\bf b} are plotted.  
{\bf d,} The ARPES data [symmetrized energy distribution curves (EDCs)] along the $k_x$ direction measured at several photon energies. All the spectra shown were accumulated at $T=15$ K.
The energy positions of spectral peaks are marked by bars and dashed curves. {\bf e,}  EDCs and the corresponding symmetrized EDCs along the $\Gamma-$L direction, measured at $k$ points marked with green circles in {\bf a}. The energy positions of dip and hump, suggesting mode-coupling, are marked by arrows and circles, respectively, on EDCs. Error bars in {\bf b} represent uncertainty in estimating the spectral peak positions.}
\label{fig1}
\end{figure*}

\newpage

\begin{figure*} \includegraphics[width=4in]{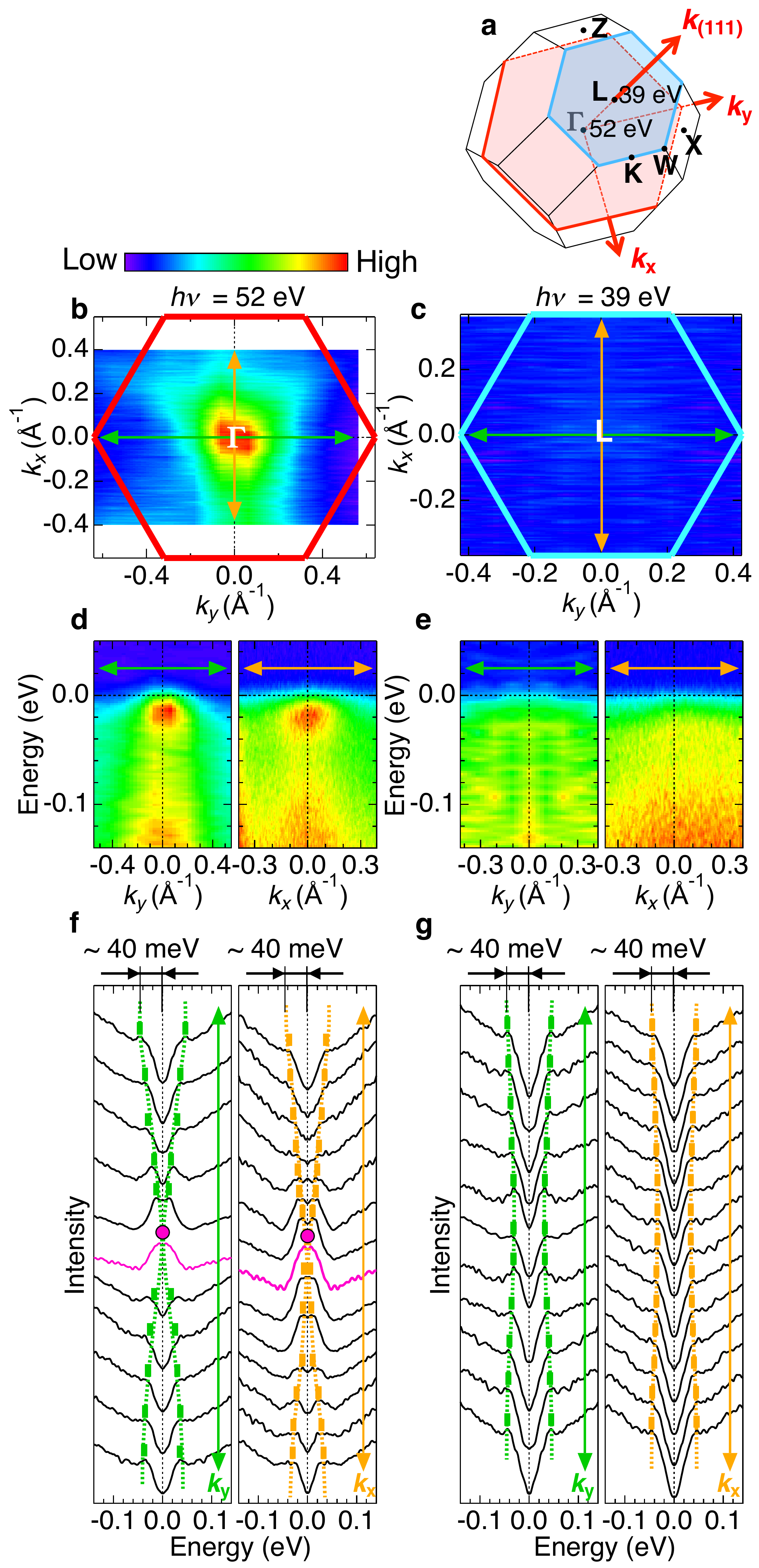} 
\caption{{\bf ARPES intensity mapping at the Fermi level across $\Gamma$ and L points.}
{\bf a,} Brillouin zone, showing  momentum sheets along which  ARPES data were measured. 
The momentum cuts at $h\nu=$ 52 eV  and 39 eV crosses  the $\Gamma$ and L points, respectively, in the 3rd Brillouin zone.
{\bf b,c,} ARPES intensities at $E_{\rm F}$ in the $k_x - k_y$ sheet crossing $\Gamma$  and L measured at $h\nu=$ 52 eV and 39 eV, respectively.  For the data at $h\nu=$39 eV, only the positive side of $k_y$ were measured, and it is reflected to the negative side.
{\bf d,} ARPES dispersion maps for $h\nu=$ 52 eV, measured along $k_y$  and $k_x$ crossing $\Gamma$ (a green and orange arrows in {\bf b}, respectively).
{\bf e,} The same data as {\bf d}, but for $h\nu=$ 39 eV, which cross the L point. 
{\bf j,g,}  EDCs of {\bf d} and {\bf e}, respectively, symmetrized about $E_{\rm F}$.  The dimension arrows estimate the band width of $\sim 40$ meV in the occupied side. }
\label{fig1}
\end{figure*}

\newpage 

\begin{figure*} \includegraphics[width=6.5in]{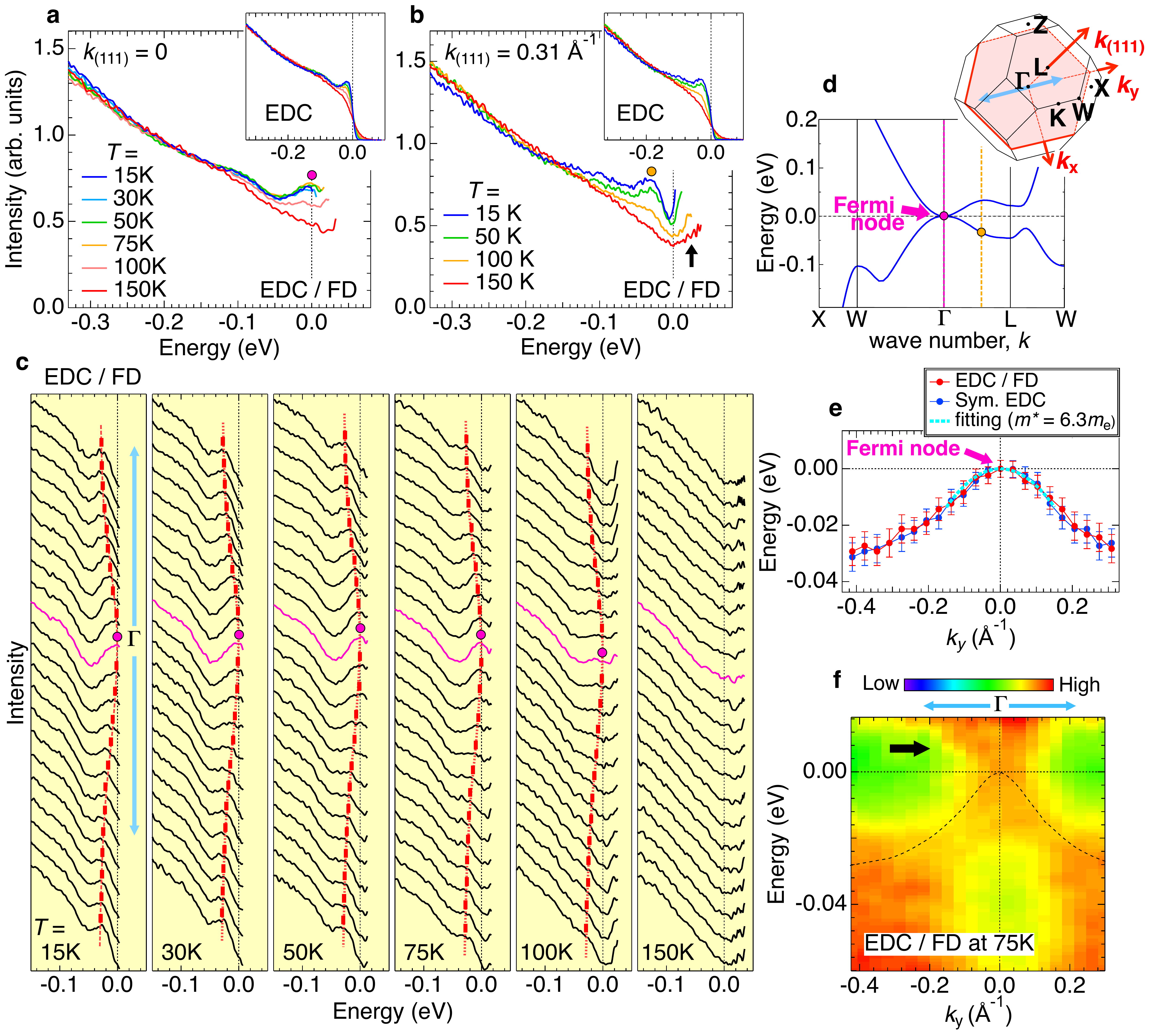} 
\caption{{\bf Temperature evolution of ARPES spectra and evidence for the Fermi nodal state.}
{\bf a,b,} Temperature variation of FD-divided EDCs  
measured at  $\Gamma$ ($h\nu=10.5$ eV) and off $\Gamma$ ($h\nu=21.2$ eV), respectively.  In  {\bf d}, the observed $k$ points for  {\bf a} and {\bf b} are marked by purple and orange dashed lines, respectively. The unoccupied side is displayed up to the energy of 3$k_BT$.  
The spectral peak positions are marked by colored circles. 
The  original EDCs used for the analysis are plotted in the insets. 
{\bf c,} The  FD-divided EDCs along a momentum cut indicated in {\bf d} with a light blue arrow measured at various temperatures. 
The spectral peak positions are indicated by red bars and dashed curves. 
The Fermi node is marked by a magenta circle. At $T=150$ K, the spectral peaks are 
strongly suppressed, and thus the dispersion can not be determined. 
{\bf d,} Band dispersions obtained by the first principles calculation, magnified in the region close to $E_{\rm F}$.  
In the inset, the Brillouin zone of Pr$_2$Ir$_2$O$_7$ is shown.
{\bf e,} The energy dispersion along a momentum cut crossing $\Gamma$ (light blue arrow in {\bf d}), 
 determined from the peak positions of the FD-divided spectra at $T = 75$K in {\bf c} (red circles).
The same plots, but determined from the  peak positions of symmetrized spectra, are superimposed (blue circles).
The quadratic dispersion (light blue dashed curve) nicely fits the data. 
{\bf f,} ARPES dispersion map crossing $\Gamma$ (light blue arrow in {\bf d}) measured at $h\nu=10.5$ eV. 
The original ARPES intensities are divided by the energy-resolution convoluted Fermi function at the measured temperature of 
$T=75$ K to remove the cut-off effect near $E_{\rm F}$.
The tick black arrow points to the remarkable spectral weight above $E_F$, consistent with the existence of conduction touching band in the unoccupied side. Error bars in {\bf e} represent uncertainty in estimating the spectral peak positions.}
\label{fig1}
\end{figure*}

\end{document}




\title{Supplementary information for\\
Quadratic Fermi Node in a 3D Strongly Correlated Semimetal}

\author{Takeshi Kondo}
\affiliation{ISSP, University of Tokyo, Kashiwa, Chiba 277-8581, Japan}

\author{M.~Nakayama} 
\affiliation{ISSP, University of Tokyo, Kashiwa, Chiba 277-8581, Japan}

\author{R.~Chen} 
\affiliation{Physics Department, University of California, Santa Barbara, California 93106, USA}
\affiliation{Physics Department, University of California, Berkeley, California 94720, USA}
\affiliation{Molecular Foundry, Lawrence Berkeley National Laboratory, Berkeley, California 94720, USA}

\author{J.J.~Ishikawa} 
\affiliation{ISSP, University of Tokyo, Kashiwa, Chiba 277-8581, Japan}

\author{E.-G.~Moon} 
\affiliation{Physics Department, University of California, Santa Barbara, California 93106, USA}
\affiliation{Department of Physics, Korea Advanced Institute of Science and Technology, Daejeon 305-701, Korea}

\author{T.~Yamamoto} 
\affiliation{ISSP, University of Tokyo, Kashiwa, Chiba 277-8581, Japan}

\author{Y.~Ota} 
\affiliation{ISSP, University of Tokyo, Kashiwa, Chiba 277-8581, Japan}

\author{W.~Malaeb} 
\affiliation{ISSP, University of Tokyo, Kashiwa, Chiba 277-8581, Japan}
\affiliation{Physics Department, Faculty of Science, Beirut Arab University, Beirut, Lebanon}

\author{H.~Kanai} 
\affiliation{ISSP, University of Tokyo, Kashiwa, Chiba 277-8581, Japan}

\author{Y.~Nakashima} 
\affiliation{ISSP, University of Tokyo, Kashiwa, Chiba 277-8581, Japan}

\author{Y.~Ishida} 
\affiliation{ISSP, University of Tokyo, Kashiwa, Chiba 277-8581, Japan}

\author{R.~Yoshida} 
\affiliation{ISSP, University of Tokyo, Kashiwa, Chiba 277-8581, Japan}

\author{H.~Yamamoto} 
\affiliation{ISSP, University of Tokyo, Kashiwa, Chiba 277-8581, Japan}

\author{M.~Matsunami} 
\affiliation{UVSOR Facility, Institute for Molecular Science, Okazaki 444-8585, Japan}
\affiliation{Energy Materials Laboratory, Toyota Technological Institute, Nagoya 468-8511, Japan}

\author{S.~Kimura} 
\affiliation{UVSOR Facility, Institute for Molecular Science, Okazaki 444-8585, Japan}
\affiliation{Graduate School of Frontier Biosciences, Osaka University, Suita, Osaka, 565-0871, Japan }

\author{N.~Inami} 
\affiliation{Institute of Materials Structure Science, High Energy Accelerator Research Organization (KEK),  Tsukuba, Ibaraki 305-0801, Japan}

\author{K.~Ono} 
\affiliation{Institute of Materials Structure Science, High Energy Accelerator Research Organization (KEK),  Tsukuba, Ibaraki 305-0801, Japan}

\author{H.~Kumigashira} 
\affiliation{Institute of Materials Structure Science, High Energy Accelerator Research Organization (KEK),  Tsukuba, Ibaraki 305-0801, Japan}

\author{S.~Nakatsuji} 
\affiliation{ISSP, University of Tokyo, Kashiwa, Chiba 277-8581, Japan}
\affiliation{PRESTO, Japan Science and Technology Agency (JST), 4-1-8 Honcho Kawaguchi, Saitama 332-0012, Japan}

\author{L.~Balents} 
\affiliation{Kavli Institute for Theoretical Physics, Santa Barbara, California 93106, USA}

\author{S.~Shin} 
\affiliation{ISSP, University of Tokyo, Kashiwa, Chiba 277-8581, Japan}

\date{\today}
\maketitle

{\bf Supplementary Note 1: Paramagnetic Band Calculations for cubic {\protect\airo} }\\
In this supplemental section, we discuss the electronic structure calculations in the paramagnetic state.  Calculations were carried out using both the standard generalized gradient approximation (GGA) \cite{gga} and the Tran-Blaha modified Becke-Johnson (TB-mBJ) \cite{TB-mBJ} exchange potential as implemented in Wien2k \cite{wien2k}, for a series of different A site ions. An RKmax parameter 7.0 was chosen and the wave functions were expanded in spherical harmonics up to $l_{\text max}^{\text wf}=10$ inside the atomic spheres and  $l_{\text max}^{\text pot}=4$ for non-muffin tins. Bulk A$_2$Ir$_2$O$_7$ (A=rare earth element) has a cubic crystal structure with the space group Fd$\bar 3$m. The experimental lattice parameters \cite{Chan-PrEustr,Yamada2012-Ndstr,shapiro2012structure, taira2001magnetic} were used for A= Pr, Nd, Eu and Y in both GGA and TB-mBJ paramagnetic calculations, and spin-orbit coupling was applied to both the heavy rare earth element and the Ir electrons. The paramagnetic GGA and TB-mBJ calculations put the 4f states of the rare earth element at the Fermi energy; to avoid this, since the 4f electrons are highly localized, their potential is shifted by a constant.

The TB-mBJ method is believed to produce improved results for small band gap systems \cite{meritlimit-mBJ}, and has been widely used in studies of topological insulators \cite{feng2010halfTI}.  Both methods yield almost indistinguishable results away from the Fermi energy, with a slight decrease of band-width in the TB-mBJ calculation, and small differences near $E_F$, as shown in Supplementary Fig. \ref{supple-fig1}.  We observed the symmetry-required nodal band touching at $\Gamma$ for all calculations, but the Fermi energy is shifted from the nodal point by an amount that decreases with increasing A (A= Y, Eu, Nd and Pr) site ionic radius in both methods \cite{Onoda}. This occurs because of accidental crossing of states near the L point, where the valence band rises and approaches the Fermi energy, especially in the smaller rare earths.  For example, a small shift of the Fermi level of about 30 meV in the GGA approximation is observed in Pr$_2$Ir$_2$O$_7$, while the shift is about 70 meV in Y$_2$Ir$_2$O$_7$.   In general, the GGA calculation underestimates the effects of correlations, which tend to narrow the bands and to push occupied states deeper below the Fermi energy.  Hence our expectation is that the accidental crossing near L, which is responsible for the Fermi level shift in GGA, will be suppressed by correlations.  The TB-mBJ method may be regarded as a crude way to do this.  Indeed, in the presumed more accurate TB-mBJ method, all the paramagnetic band structures for the A$_2$Ir$_2$O$_7$ series show smaller shifts of the Fermi level compared to their GGA counterparts. In the case of A=Pr, we find the shift vanishes and the nodal point occurs precisely at the Fermi energy, and a shift value of around 7 meV is observed in the case of A=Y. While there is a universal trend of smaller shift of the Fermi energy as rare earth ionic radius increases for the rare earth elements we have tested, the small differences between GGA and TB-mBJ suggest that some theoretical uncertainty remains.  We note, however, that we expect methods that include correlations more accurately, such as LDA+DMFT, are likely to further suppress the states near the Fermi energy at L even beyond TB-mBJ, favoring placing the Fermi level precisely at the node.
 
The above calculation treats Pr as non-magnetic by artificially shifting the levels away from the Fermi energy.  More properly, these levels are highly localized and, due to large Coulomb repulsion, form local moments, which fluctuate thermally in the range of the ARPES measurements.  A calculation which includes the localized but thermally fluctuating paramagnet state of the Pr spins requires Dynamical Mean Field Theory (DMFT), which is beyond the scope of this paper.  However, we expect that the Pr moments are in any case so strongly localized that they couple only weakly to the Ir electrons.  We can test this, and thereby gauge the magnitude of such effects, by studying the system at low temperature using the GGA+U technique.  This treats the localization of the 4f moments but does not allow them to fluctuate.  It has indeed been inferred that in Pr$_2$Ir$_2$O$_7$, the Pr moments becomes quasi-static and form some 2-in/2-out structure below 0.3K.  

As the 2-in/2-out states are degenerate, we instead focus on the simpler all-in/all-out configuration of 4f moments.  Although this is not the proper arrangement in Pr$_2$Ir$_2$O$_7$, it still shows the magnitude of the effects to be expected from Pr magnetic order, and specifically from time-reversal symmetry breaking. To include the non-collinear magnetism, the calculation is performed in WIENNCM (non-collinear magnetic version of the Wien2k package) \cite{wienncm_PRB}. We take U on the $4f$ electrons to be 6eV and U on the Ir $5d$ electrons to be 0eV. It is inevitable that the Ir $5d$ electrons acquire a small magnetic moment, through their interaction with the Pr ones.   The largest Ir moment introduced in the GGA+U calculation is below 0.004$\mu_{B}$. Supplementary Fig.~\ref{supple-fig-Prf} shows the magnetic band structure of Pr$_2$Ir$_2$O$_7$.  Generally the structure of the Ir bands is very similar to that in the paramagnetic GGA calculation.  This proves the weak coupling of Pr moments to the Ir electrons.  However, there is indeed, as expected by symmetry, a small energy splitting of the bands in the GGA+U band structure, comparing to the originally degenerate bands in the GGA calculation. This is consistent with broken time reversal symmetry, which, as expected from low energy $k\cdot p$ theory, leads to the appearance of Weyl points along the $\Gamma-L$ branch.  The point appears near the $\Gamma$ point, along the $\Gamma$--L line, as shown in Supplementary Fig.~\ref{supple-fig-Prf}.  Note that the band bending between the L and W points is larger in the GGA+U calculation than in the TB-mBJ one, so a Fermi pocket appears in this calculation in this region in the GGA+U result.
\\

{\bf Supplementary Note 2: Energy distribution curves (EDCs) used for the symmetrization analysis}\\
In Fig. 3 of the main paper, we show the energy distribution curves (EDCs) symmetrized about $E_{\rm F}$ to remove the cut-off effect by Fermi function. Here we present the row EDCs used for the analysis  (Supplementary Fig.~\ref{EDC}). Moreover, we show some data which are discussed but not presented in the main paper (Supplementary Fig.~\ref{EDC2} and \ref{3rd_kz}).

Supplementary Fig.~\ref{EDC}b shows the EDCs measured at various photon energies from $h\nu$ = 7 eV to 18 eV. The measured momentum cuts (dashed lines in Supplementary Fig.~\ref{EDC}a) are located in the $k_x-k_{(111)}$ plane crossing $\Gamma$, L, and K. 
 As marked by magenta arrows, the quasiparticle peaks clearly disperse along $k_x$,  and that at $k_y=0$ (green curve in each panel) most approaches $E_{\rm F}$ at $h\nu$ = 10 eV as examined in the right panel of Supplementary Fig.~\ref{EDC}b. The symmetrized EDCs of Supplementary Fig.~\ref{EDC}b are plotted in Supplementary Fig.~\ref{EDC}c. (The same data at $h\nu$ = 7, 8, 9, and 10 eV are shown in Figs. 3d of the main paper.) It is clearly seen that the quadratic dispersion approaches $E_{\rm F}$ with increasing the photon energy, touches $E_{\rm F}$ at 10 eV,  and  move away from $E_{\rm F}$ again with a further increase of $h\nu$. 
This behavior is more clearly demonstrated in the right panel of Supplementary Fig.~\ref{EDC}c, where the curves at $k_x=0$ are extracted. 
The corresponding ARPES images symmetrized at $E_F$ are exhibited in the bottom panels of Supplementary Fig.~\ref{EDC}d. 
The upper panels plot the second derivatives of those.  The systematic variation of dispersions with photon energy, including the Fermi node state at $h\nu$ = 10 eV, are clearly seen. 

We have further confirmed the existence of Fermi node by observing a different momentum sheet crossing $\Gamma$, L, and W points using  another piece of sample. The $k_x-k_{(111)}$ plane and  measured momentum cuts corresponding to used photon energies (9.5 eV to 11.5 eV) are indicated in Supplementary Fig.~\ref{EDC2}a. Supplementary Fig.~\ref{EDC}b shows the EDCs along $k_x$ obtained at different photon energies.
The $h\nu$ value was swept by a fine step ($\le 0.5$ eV, see dashed lines in Supplementary Fig.~\ref{EDC2}a). The quasiparticle peak disperses along $k_x$, and  most approaches $E_{\rm F}$ at $\sim$ 10 eV as examined in the right panel of Supplementary Fig.~\ref{EDC2}b. 
The corresponding symmetrized EDCs are plotted in  Supplementary Fig.~\ref{EDC2}c.  We identify the band touching to $E_{\rm F}$ in the data at $h\nu=10.5$ eV. This is more clearly presented in the right panel of Supplementary Fig.~\ref{EDC2}c, where the curves at $k_x=0$ are extracted. The broad shape of quasiparticle peak in Pr$_2$Ir$_2$O$_7$ does not allow us to determine the exact photon energy for the $\Gamma$ point. Nevertheless, our data signify that the Pr$_2$Ir$_2$O$_7$ has a 3D band structure with a Fermi node theoretically predicted. 

In order to further validate the 3D nature of band structure in Pr$_2$Ir$_2$O$_7$, we examined it with higher photon energies ($39 \leqslant h\nu  \leqslant 60$ eV), which reach the 3rd Brillouin Zone in the $k_{(111)}$ direction.  
Supplementary Fig.~\ref{3rd_kz}b plots the EDCs at various photon energies corresponding to momentum points marked by green circles in Supplementary Fig.~\ref{3rd_kz}a. The curves are symmetrized in Supplementary Fig.~\ref{3rd_kz}c to judge whether or not the dispersion touches  $E_{\rm F}$. 
A gap clearly observed at 40 eV gets smaller with increasing $h\nu$, 
and it becomes negligible at 52 eV, where a broad single peak is obtained. With a further increase of $h\nu$, the gap opens again showing two peak structure in the spectra, being consistent with the Fermi nodal state. The energy dispersion along $k_{(111)}$ 
is consistent with that seen in the 1st Brillouin Zone observed with lower photon energies. The periodic trend over a wide range of photon energy ensures that our data provide the intrinsic 3D bulk band structure of Pr$_2$Ir$_2$O$_7$. 

In the right- and left-side panels of Supplementary Fig.~\ref{3rd_BZ}, we show the band dispersion along two $k_x-k_y$ planes across $\Gamma$ ($h\nu=52$ eV) and L ($h\nu=39$ eV), respectively. The ARPES intensities at $E_{\rm F}$ (top panels), the dispersion maps along $k_x$ and $k_y$ (middle panels), and the corresponding symmetrized EDCs (bottom panels) are also presented in Fig. 4 of the main paper. Here we plot the original EDCs before the symmetrization  (left panels in Supplementary Figs.~\ref{3rd_BZ}h-\ref{3rd_BZ}k). 
As marked by magenta arrows in each panel, the quasiparticle peaks are  observed.  Importantly, while the dispersion is clearly seen across the $\Gamma$ point (Supplementary Fig.~\ref{3rd_BZ}h and \ref{3rd_BZ}i), it is negligible across the L point (Supplementary Fig.~\ref{3rd_BZ}j and \ref{3rd_BZ}k), which agrees to the Fermi node state.   
\\

{\bf Supplementary Note 3: Agreement between the results obtained with symmetrized spectra and FD-divided spectra}\\
In the main paper, we use two different techniques to remove the Fermi cut-off effect from the ARPES spectra (EDCs). 
One is the symmetrization method, which flips the EDC about $E_F$ and add it to the original curve. 
This technique is widely used to visualize the opening of a gap, and also to determine the Fermi crossing point ($k_F$ point). 
The method is easy to use and applicable even for the low temperature spectra with negligible intensity above $E_F$, thus 
it has been commonly used for a gap estimation in the superconducting and density-wave materials. 
A weakness of this method is that it requires the particle-hole symmetry, which is not for sure in the unknown materials like Pr$_2$Ir$_2$O$_7$.   
The more straightforward way of eliminating the Fermi cut-off effect would be to divide the EDCs by the Fermi function (FD) at measured temperature. 
For the realistic analysis, the function actually used is convoluted with the experimental energy resolution. 
However, the FD-division method, though conceptually straightforward, 
also has a difficulty; the high temperature data are required, since it utilizes
the thermally populated intensity above $E_F$. Moreover, 
 the high quality data with a high signal/noise ratio are essential in order to reliably reproduce 
 the spectral shape close to $E_F$. 
Therefore, the best way to extract the band dispersion close to $E_F$ is to 
 apply both the analyses of the symmetrization and FD-division methods to the same data, and confirm the consistency between the two results. 
 
 In Supplementary Fig.\ref{EDCvsSym}, we demonstrate that the results obtained with these two methods agree with each other,
 pointing to the realization of the quadratic Fermi nodal state in Pr$_2$Ir$_2$O$_7$. 
Supplementary Figure \ref{EDCvsSym}a plot the EDCs along a momentum cut across $\Gamma$ (light blue arrow in Supplementary Fig.\ref{EDCvsSym}d).
Here we use the data at $T=75$K, which fulfill the following two conditions: the surviving of quasiparticles  
and the sufficient intensity of thermally populated spectral weight above $E_F$. 
The EDCs in Supplementary Fig.\ref{EDCvsSym}a 
divided by the Fermi function and symmetrized about $E_F$ are 
exhibited in Supplementary Figs. \ref{EDCvsSym}b and \ref{EDCvsSym}c, respectively. 
The effect of Fermi cut-off near $E_F$
is removed after these treatments, and importantly, the dispersion of the spectral peaks (color bars and dashed curves) 
touches $E_F$ at the $\Gamma$ point (magenta curves) in both the cases. 
The results are more clearly demonstrated in Supplementary Fig. \ref{EDCvsSym}f, where the peak positions in Supplementary Figs. \ref{EDCvsSym}b and \ref{EDCvsSym}c
are extracted. We find that the plots obtained by the symmetrization and the FD-division techniques (red and blue circles, respectively) almost
perfectly matches with each other, both showing the Fermi node state. 
The consistency between the two different analyses strongly support our conclusion. 
A further support is presented in Supplementary Fig. \ref{EDCvsSym}e, which plot the 
the ARPES image after the FD-division treatment. We find the remarkable intensities above $E_F$ (pointed by a tick black arrow), which implies   
the existence of the conduction band touching on the unoccupied side, as expected in the band calculation. 
\\

{\bf Supplementary Note 4: Comparison of  ARPES data with  band calculations: agreement with the TB-mBJ calculation, and disagreement with the GGA calculation}\\
The relevance of band calculation could be judged by directly comparing it with the experimental results. The ARPES is the most direct technique to determine the momentum-resolved band structure,  thus it allows us to achieve this aim. Here we carefully compare the ARPES data and  two  different band calculations of GGA and TB-mBJ, and demonstrate that the TB-mBJ method more correctly reproduces the real band structure in Pr$_2$Ir$_2$O$_7$. 

There are mainly two differences between the results of the GGA and TB-mBJ calculations as summarized in the left and middle panels in Supplementary Fig. \ref{GGA_vs_TB-mBJ}. The first difference is seen in the energy dispersion from $\Gamma$ toward L point: in GGA, it approaches $E_F$ (red arrow in Supplementary Fig. \ref{GGA_vs_TB-mBJ}b), whereas goes toward higher binding energies in the TB-mBJ (red arrow in Supplementary Fig. \ref{GGA_vs_TB-mBJ}e). Secondly, the two calculations are distinguished by whether or not the Fermi pocket exists around the L point: the GGA has the doughnut-shaped Fermi pockets surrounding the L point (see Supplementary Fig. \ref{GGA_vs_TB-mBJ}a and red dashed circle in \ref{GGA_vs_TB-mBJ}b), whereas there is no Fermi crossings in the result of TB-mBJ with fully gapped along the L-W  (Supplementary Fig. \ref{GGA_vs_TB-mBJ}e). 

We summarize our ARPES results in the right-side panels of Supplementary Fig. \ref{GGA_vs_TB-mBJ}. We find that the band structure disperses toward higher binding energies from the $\Gamma$ towards L point in a monotonic fashion (red arrow in Supplementary Fig. \ref{GGA_vs_TB-mBJ}h). Secondly, a gap keeps opening all the way from $\Gamma$ to W point with no indication of Fermi crossings (or Fermi pockets). These two   
features are reproduced by the TB-mBJ calculation, while disagree with the GGA calculation. 
The TB-mBJ is thus more relevant than the GGA  for the calculation of electronic system in Pr$_2$Ir$_2$O$_7$.
Most importantly, the absence of pockets off the $\Gamma$ point requires the touching of quadratic valence and conduction band just at $E_F$ to hold the charge neutrality, which justifies our conclusion. 

Nonetheless, we also find a clear difference between the ARPES data and the TB-mBJ calculation. 
The  ARPES results exhibit a much narrower band width of $\sim 40$meV in the occupied side than the calculation ($>$100 meV). 
Intriguingly, the band width of Pr$_2$Ir$_2$O$_7$ is thus more than one order of magnitude narrower than the band structure observed in HgTe \cite{HgTe_ARPES}. It is actually consistent with the view that the Pr$_2$Ir$_2$O$_7$ is a strongly correlated analog of HgTe with a mathematically identical quadratic node at the Fermi energy. 
\\

{\bf Supplementary Note 5: Dramatic peak suppression at elevated temperatures and aging check after the temperature sweeping}\\
In Fig. 5 of the main text, we investigate the temperature evolution of ARPES spectra measured, and demonstrate a strong suppression of the spectral peak at elevated temperatures. 
Here we confirm that this is an intrinsic property of Pr$_2$Ir$_2$O$_7$ rather than an aging effect on the sample surface.

Supplementary Fig.~\ref{aging}b and \ref{aging}a show EDCs and the same curves symmetrized about $E_{\rm F}$ over a wide range of temperature from 15K up to 300K. (The same data up to 150K are presented in the main paper.) We used 21.2 eV photons (He discharge lamp), which locate the  $k$ point  in  nearly the midpoint between $\Gamma$ and L (a green circle in Supplementary Fig. \ref{aging}c and a green dashed line in the inset  of Supplementary Fig. \ref{aging}d). The upturn behavior toward higher energies beyond $E_{\rm F}$ (arrow in Supplementary Fig. \ref{aging}a) is  visible regardless of temperature, which is consistent with the theoretical prediction that the quadratic band similar to that in the occupied side is stayed in the unoccupied side (see inset of Supplementary Fig. \ref{aging}d).  In order to check where or not there is an aging effect on the spectra, 
we cooled the sample down to the lowest temperature after the temperature scan, and compare  the spectral shapes  before and after the measurement, as shown in Supplementary Fig. \ref{aging}d.
The spectral shape  just after the sample cleaving is almost perfectly reproduced, which validates that the remarkable variation in spectral shape with temperature is an intrinsic feature of Pr$_2$Ir$_2$O$_7$, mostly due to the  strong electron correlation.
\\

{\bf Supplementary Note 6: Signature of the unoccupied conduction band observed by a laser ARPES with an ultra-high energy resolution}\\
The ARPES is the technique to measure the occupied side of band dispersion. However, the energy states slightly above $E_F$ can 
be evaluated by examining the thermally populated spectral weight divided by the Fermi function (FD).
The electronic system of Pr$_2$Ir$_2$O$_7$ is strongly correlated, and the quasiparticle peaks vanish at elevated temperatures.
This feature prevents one from a definitive determination of the unoccupied band dispersion with this technique. 
Nevertheless, we have obtained the upturn behavior in the spectra above $E_F$ (Fig. 5b and Supplementary Fig.\ref{aging}a),  signifying the existence of the conduction band in the unoccupied side. 
To verify it further, here we apply the FD-division technique to the data measured by a 7-eV laser ARPES, which is capable of an ultra-high energy resolution ($\sim$2meV), bulk sensitive (a long photoelectron escape depth: $\sim 100 \rm {\AA}$), and high signal/noise ratio measurements. 
These performances has huge advantages in reproducing the intrinsic spectral shape above $E_F$ by the FD-devision.  
The $k_z$ position corresponding to 7eV photons is located at the midpoint between $\Gamma$ and L, where the unoccupied conduction band is predicted to stay at almost the same energy distance from $E_F$ as the occupied valence band (Supplementary Fig.\ref{LaserARPES}e). Therefore, the signature for the conduction band is expected to be obtained from the spectral tail above $E_F$ at high temperatures. 

First, we have tested the FD-division technique to the spectra of the evaporated Au, which is electrically contacted to the samples. 
 We confirmed, in Supplementary Fig.\ref{LaserARPES}a, that the Au spectra divided by the energy-resolution convoluted FD becomes flat at least up to $3k_BT$ above $E_F$, meaning that  
 the  analysis we use effectively removes the Fermi cut-off effect near $E_F$.  
 We have accumulated the spectra of Pr$_2$Ir$_2$O$_7$ with  exactly  the same   experimental setting as that for the Au measurements,
 and also applied the identical FD-division analysis to the data in the Supplementary Fig.\ref{LaserARPES}b. 
 We have reproduced the previous results of spectral peak suppression at elevated temperatures, indicating
 the significance of strong correlation in this material. 
Furthermore, the upturn behavior toward higher energies beyond $E_F$ is also obtained in the present data (see the bottom panel of Supplementary Fig.\ref{LaserARPES}b). 
Here we emphasize that the thermally populated spectral weight above $E_F$ seen in the data is large enough to reliably extract the intrinsic electronic state in the unoccupied side by the FD-division technique (see Supplementary Fig.\ref{LaserARPES}d) at least up to $3k_BT$ above $E_F$.  
This feature is exactly expected when the conduction touching-band exists in the unoccupied side.
 The consistent results from the data sets at different photon energies justify our conclusion. 
 
After the above measurement of temperature scan, we cooled the sample again down to the lowest temperature 
and compared the spectral shape with that taken just after the sample cleaving.
As demonstrated in Supplementary Fig.\ref{LaserARPES}c, the two curves are almost perfectly overlapped with each other, meaning that the sample aging is negligible. 
This assures that the strong suppression of spectral peaks at elevated temperatures seen in  Supplementary Fig.\ref{LaserARPES}b
is an intrinsic phenomenon for Pr$_2$Ir$_2$O$_7$. 
\\

{\bf Supplementary Note 7: First Principle Calculations for {\protect\priro} with Uniaxial Strain}\\
The TB-mBJ paramagnetic band structure calculations show Pr$_2$Ir$_2$O$_7$ has a nodal Fermi point, at which the doubly degenerate, quadratically dispersing conduction and valence bands touch at the zone center, right at the Fermi level.   This suggests that Pr$_2$Ir$_2$O$_7$ is very sensitive to perturbations, such as time reversal symmetry or cubic symmetry breaking terms, giving rise to the possibility of various novel phases.  Indeed, the effect of small uniaxial strain near the $\Gamma$ point can be deduced, up to a sign, by symmetry within the effective mass method \cite{Leon_PRL2013}.   For the proper sign, uniaxial strain is predicted from this approach to open a gap and produce a topological insulator.  However, on pure symmetry grounds, it is not clear whether uniaxial compression or expansion is appropriate.

In this supplemental section, we consider this problem microscopically.  Uniaxial strain is applied along the $\langle 111\rangle$ direction, which breaks the cubic symmetry and converts the system into a rhombohedral crystal structure with space group R$\bar 3$m. Since TB-mBJ has no exchange functional \cite{TB-mBJ}, the GGA functional is used for structural optimization, both for the lattice parameters and the atomic positions. In the hexagonal unit cell setting, the in-plane lattice parameter is fixed and the out-of-plane lattice parameter is varied to minimize the total energy. We consider the case of $1\%$ ($a=$1.01$a_0$, $a_0=7.35$ \AA, the bulk in-plane lattice parameter in the hexagonal cell) and $5\%$ ($a=$1.05$a_0$) in-plane tensile strain, which is equivalent to uniaxial compressive pressure along $\langle 111\rangle$ direction, 
in Supplementary Figs. \ref{supple-fig2}c and \ref{supple-fig2}d, respectively, 
and $5\%$ ($a=$0.95$a_0$) in-plane compressive strain in Supplementary Fig. \ref{supple-fig2}a.
 Using the relaxed structure, the TB-mBJ paramagnetic band structure calculation shows that a gap opens with in-plane tensile strain, and the gap increases as the strain intensifies, as shown in Supplementary Figs. \ref{supple-fig2}c and \ref{supple-fig2}d.  The quadratic band touching state at the $\Gamma$ point is very sensitive to the cubic symmetry breaking perturbation, and immediately the degeneracy at the $\Gamma$ point is split as strain is applied.  We indeed observe the direct gap at the $\Gamma$ point is substantial and grows proportionally with strain.  However, there is a closing of different states away from $\Gamma$, closer to the Z point ($\pi$,$\pi$,$\pi$), which results in a reduced total gap. A small gap, about 4 meV, exists in the case of 1 $\%$ tensile strain (Supplementary Fig. \ref{supple-fig2}c) and the gap is 60 meV in the case of 5 $\%$ strain (Supplementary Fig. \ref{supple-fig2}d). The GGA calculation shows metallic state in the 1 $\%$ case and a much reduced gap, 8 meV, with the 5$\%$ tensile strain, which is attributed to the well-known fact that GGA and LDA greatly underestimated the gap.  On the other hand, by applying in-plane compressive strain (Supplementary Fig. \ref{supple-fig2}a), Pr$_2$Ir$_2$O$_7$ becomes metallic and shows a band crossing along the $\langle 111\rangle$ direction. This band crossing is indeed expected from the effective mass theory of the perturbed Fermi node, for the strain of sign opposite to that which produces the topological gap \cite{Leon_PRL2013}.

To confirm that the gapped state produced by uniaxial pressure along $\langle 111\rangle$ direction is topological, we further examine the Z$_2$ topological invariant. Since the strained structure preserves inversion symmetry, the Z$_2$ invariant is expressed by \cite{TI-inv,MoorePRB,RevModPhys.82.3045}

\begin{equation}
(-1)^\nu=\prod_i \delta_i, 
\end{equation}
where $\delta_i=\prod_{m=1}^{N} \xi_{2m}(\Gamma_i)$ is the parity eigenvalue at the time-reversal invariant momentum $i$ (TRIM), which are tabulated in Supplementary Table 1. The Z$_2$ class is (1;000), which verifies that Pr$_2$Ir$_2$O$_7$ under uniaxial pressure along $\langle 111\rangle$ direction indeed turns into a strong topological insulator with strong correlelation.  Note that the parity eigenvalues at all points except the zone center are essentially already defined in the cubic structure, so that the  topological nature is already incipient without strain.

\begin{table}
\setlength{\tabcolsep}{12pt}
\begin{tabular}{c| c| c|c }
\hline  \hline
$\Gamma$ & L & Z & F \\ 
\hline 
+1 & -1 & -1 & -1 \\ 
\hline \hline
\end{tabular} 
\label{table:hop}
\caption{Parity eigenvalues at the TRIM for Pr$_2$Ir$_2$O$_7$ under uniaxial pressure along $\langle 111\rangle$. }
\end{table}

We also consider the surface spectrum of the strained material for further evidence for the topological insulating state.  In the latter, an odd number of gapless surface Dirac cones are expected, and we indeed find them, as we now show.  To locate the surface state, we first obtain the tight binding Hamiltonian for $\langle111\rangle$ strained Pr$_2$Ir$_2$O$_7$ using maximally localized Wannier functions\cite{wien2wannier,wannier90}, in the $J_{\text{eff}}=1/2$ basis. Both the orbital and spinor parts are rotated to the local axes of each inequivalent Ir atom. The rhombohedral unit cell is modified to the hexgonal unit cell, and we consider a thick slab of Pr$_2$Ir$_2$O$_7$ to accommodate with the surface state. The slab is fixed to include 73 Ir layers, with identical Ir kagome layers on the upper and lower surface of the slab.  Supplementary Fig.~\ref{supple-fig4}a shows the band structure of the bulk and the surface state for Pr$_2$Ir$_2$O$_7$ under 5\% uniaxial strain on the $\langle 111\rangle$ surface, choosing the large strain to better  distinguish bulk and surface states. The Dirac points are located at the $\overline{\textrm{M}}$ points. On the surface Brillouin zone there are four TRIM, which are located at $\overline{\Gamma}$ and three $\overline{\textrm{M}}$ points, which are equivalent by 120$^\circ$ rotations. There are three symmetry operations of the crystal with space group R$\bar 3$m, including the inversion, 3-fold rotation along the $\langle 111\rangle$ direction and two-fold rotation around $x$ direction (orthorgonal to $\langle 111\rangle$). The 2D Brillouin zone also obeys these symmetry operations, making the three $\overline{\textrm{M}}$ points all equivalent. In total there are three Dirac cones around the TRIM on the 2D surface, which is consistent with the parity argument.

\clearpage 

\begin{figure*}
\includegraphics[width=5in]{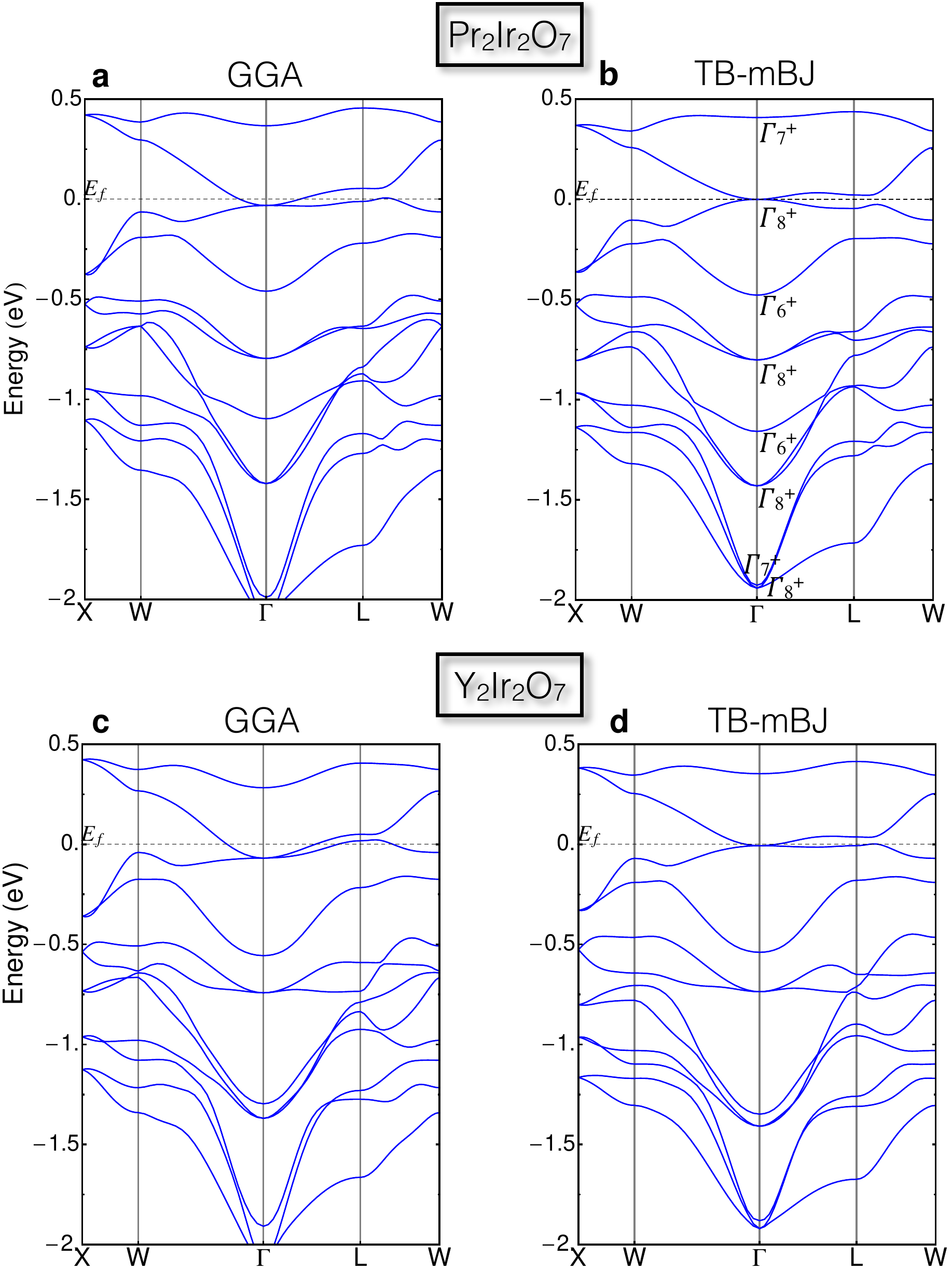}
\caption{{\bf a,b,} Paramagnetic band structure within GGA and TB-mBJ, respectively,  in the case of A=Pr. In $\bf b$, the bands are identical to Fig. 2e, with irreducible representation group at the $\Gamma$ point shown. The main band characters around the Fermi energy are of Ir $d$ bands.  {\bf c,d,} Paramagnetic band structure within GGA and TB-mBJ, respectively, in the case of A=Y. 
 }  
\label{supple-fig1}
\end{figure*}
\newpage

\begin{figure*}
\includegraphics[width=4in]{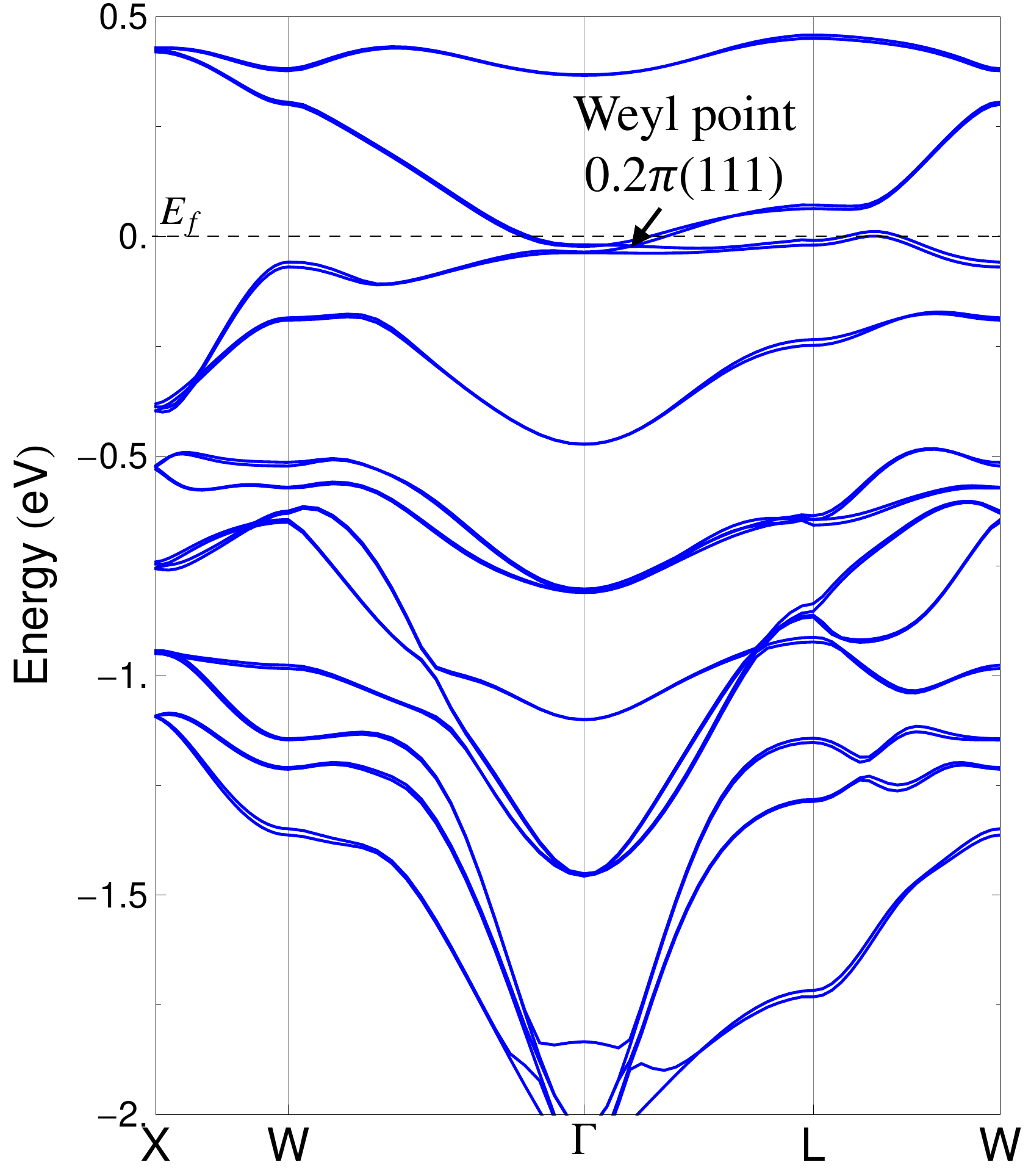}
\caption{The magnetic band structure calculation on Pr$_2$Ir$_2$O$_7$, with Pr $4f$ moment pointing along the all-in all-out direction.}  
\label{supple-fig-Prf}
\end{figure*}
\newpage 

\begin{figure*}  
\includegraphics[width=5.7in]{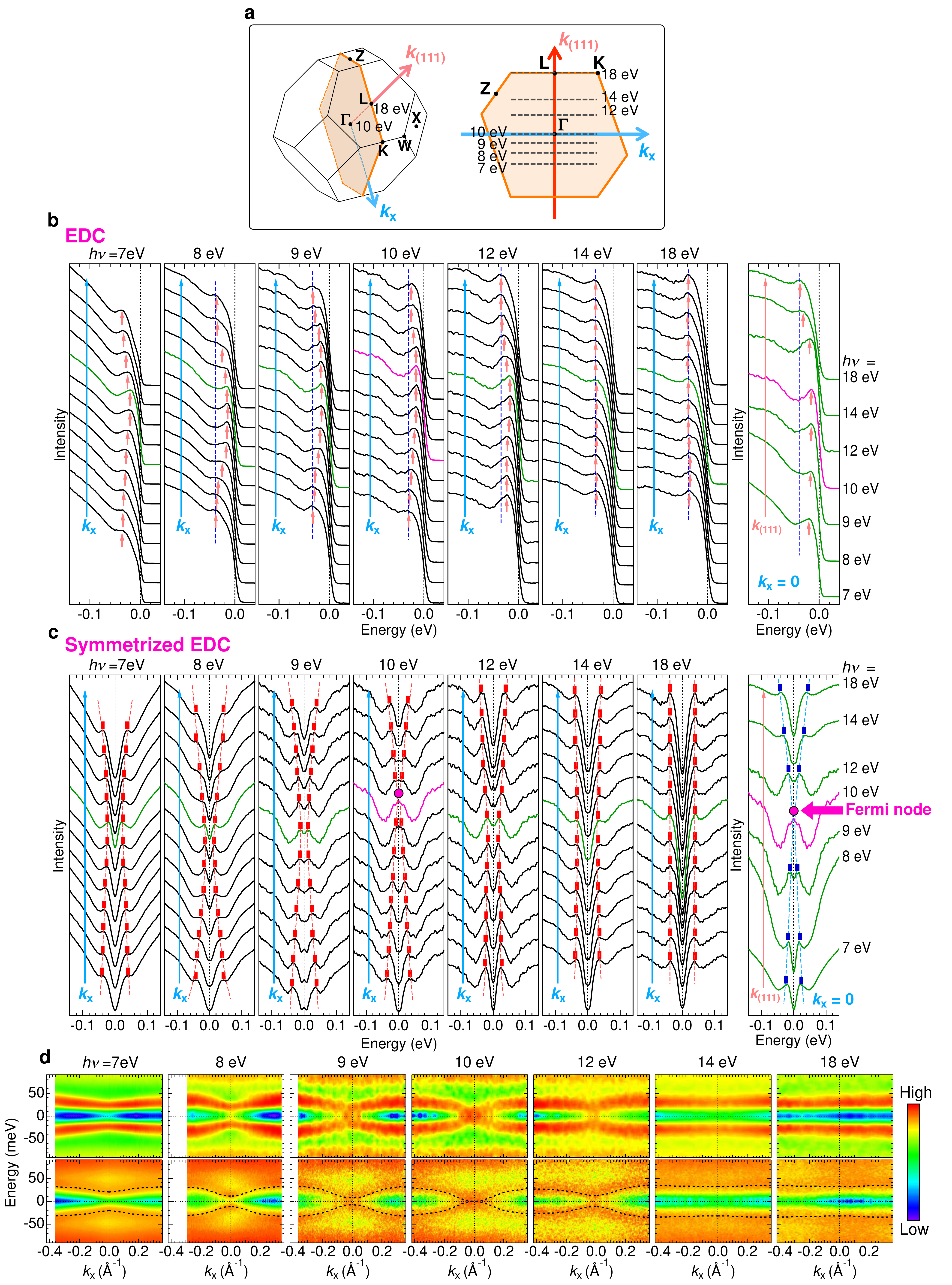}
\caption{$\bf a,$  Brillouin zone, indicating the measured momentum plane  across $\Gamma$, L and K (painted by orange) and momentum cuts for various photon energies (dashed lines). 
$\bf b,$ Energy distribution curves (EDCs) measured by ARPES along $k_x$ (defined in $\bf a$). 
The used photon energy is described on the top of each panel, and the corresponding $k_{(111)}$ location 
is indicated in $\bf a$. In the right panel (green and magenta curves), the EDCs at $k_x=0$ are extracted.  
The magenta arrows mark the energy position of spectral peak. A blue dashed line is plotted in each panel to confirm an energy dispersion of the peaks. 
$\bf c,$ The symmetrized EDCs of $\bf b$. The red bars and dashed curves indicate the spectral peaks and the energy dispersion, respectively. In the right panel (green and magenta curves), the curves at $k_x=0$ are extracted. $\bf d,$ ARPES images symmetrized about $E_F$ obtained at various photon energies (bottom panels), 
and the second derivative of each image (top panels).  The measured momentum cuts are indicated in $\bf a$ with dashed lines.}
\label{EDC}
\end{figure*}
\newpage 

\begin{figure*}  
\includegraphics[width=6in]{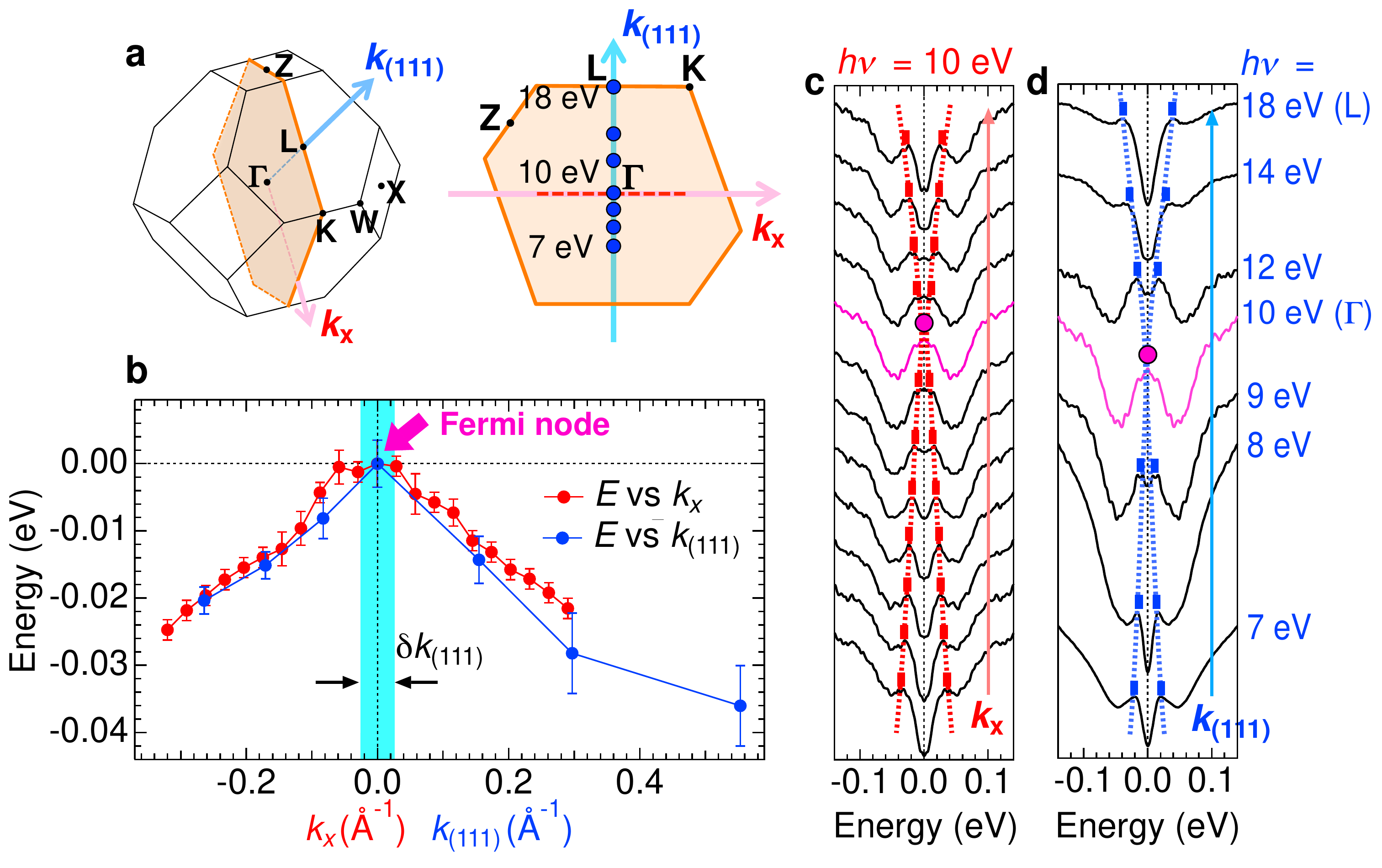}
\caption{Comparison between the energy dispersion along two directions [$k_{\rm (111)}$ and perpendicular to it ($k_x$)], and an estimation of the $k_z$-broadening ($\delta {k_z}$ or $\delta {k_{\rm(111)}}$).
  $\bf a,$ Brillouin zone of Pr$_2$Ir$_2$O$_7$. 
  $\bf b,$ The energy dispersions along the $k_x$ and $k_{\rm (111)}$ directions (red and blue circles, respectively), determined from the peak positions of spectra in $\bf c$ and $\bf d$.
  The former is measured at $h\nu $=10eV (red dashed line in the right panel of $\bf a$),
  and the latter at various photon energies (blue circles in the right panel of $\bf a$). 
The experimental $k_z$ broadening ($\delta {k_z}$ or $\delta {k_{\rm(111)}}$) is estimated to be 
$\sim 5 \%$ of Brillouin zone at $h\nu $=10eV ($\Gamma$ point), which is indicated with a light blue belt and black dimension arrows.  
$\bf c,d$ The ARPES spectra (symmetrized EDCs) along $k_x$ and $k_{\rm (111)}$, respectively. Error bars in {\bf b} represent uncertainty in estimating the spectral peak positions. 
}
\label{kz_broadening}
\end{figure*}
\newpage 

\begin{figure*}  
\includegraphics[width=5.5in]{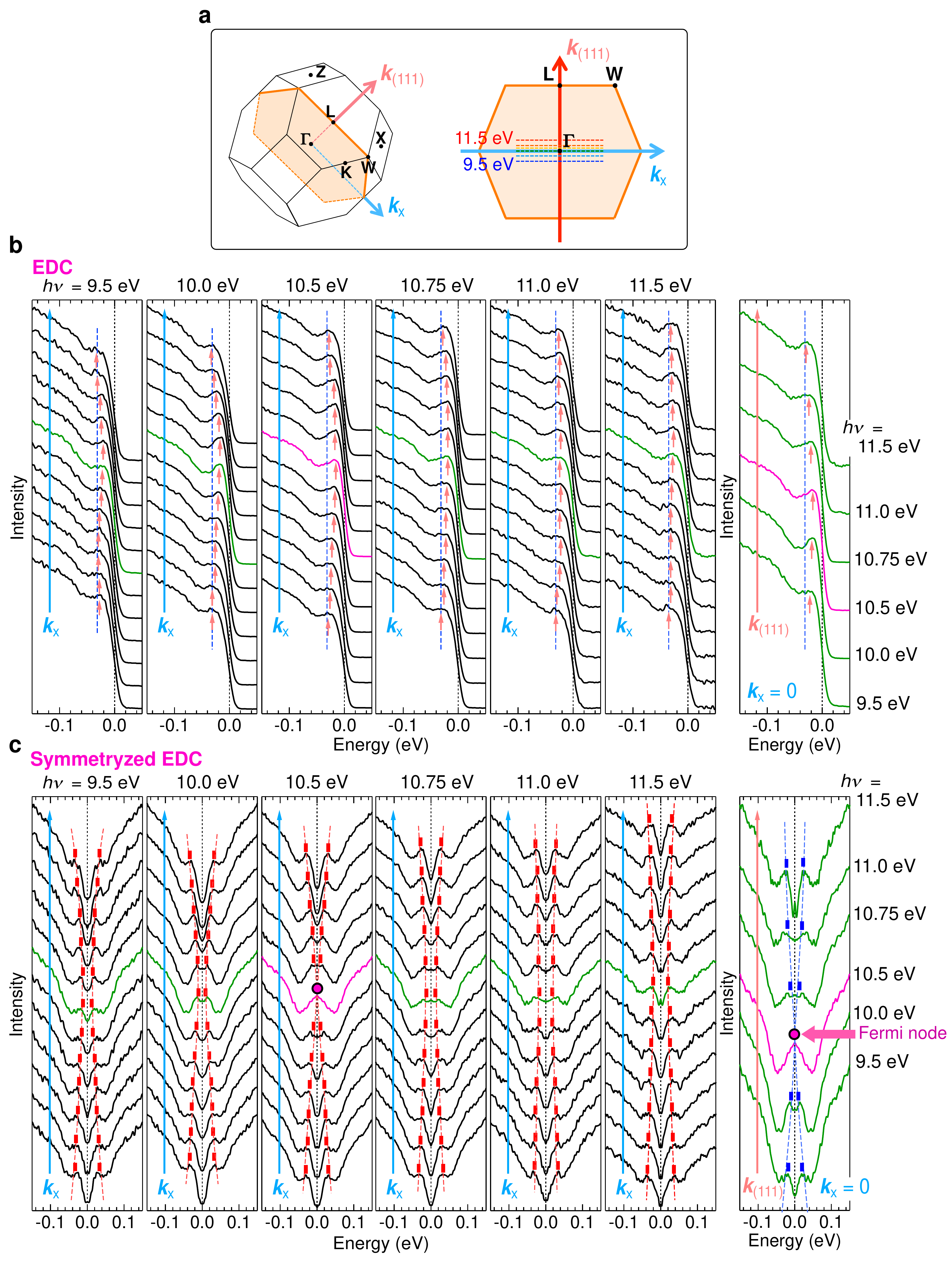}
\caption{$\bf a,$  Brillouin zone, indicating the measured momentum plane across $\Gamma$, L and W (painted by orange) and momentum cuts for various photon energies (dashed lines). 
$\bf b,$ Energy distribution curves (EDCs) measured by ARPES along $k_x$ (defined in $\bf a$). 
The used photon energy is described on the top of each panel, and the corresponding $k_{(111)}$ location 
is indicated in $\bf a$. In the right panel (green and magenta curves), the EDCs at $k_x=0$ are extracted.  
The magenta arrows mark the energy position of spectral peak. A blue dashed line is plotted in each panel to confirm an energy dispersion of the peaks. $\bf c,$ The symmetrized EDCs of $\bf b$. The red bars and dashed curves indicate the spectral peaks and the energy dispersion, respectively. In the right panel (green and magenta curves), the curves at $k_x=0$ are extracted.}
\label{EDC2}
\end{figure*}
\newpage 

\begin{figure*}  
\includegraphics[width=4.5in]{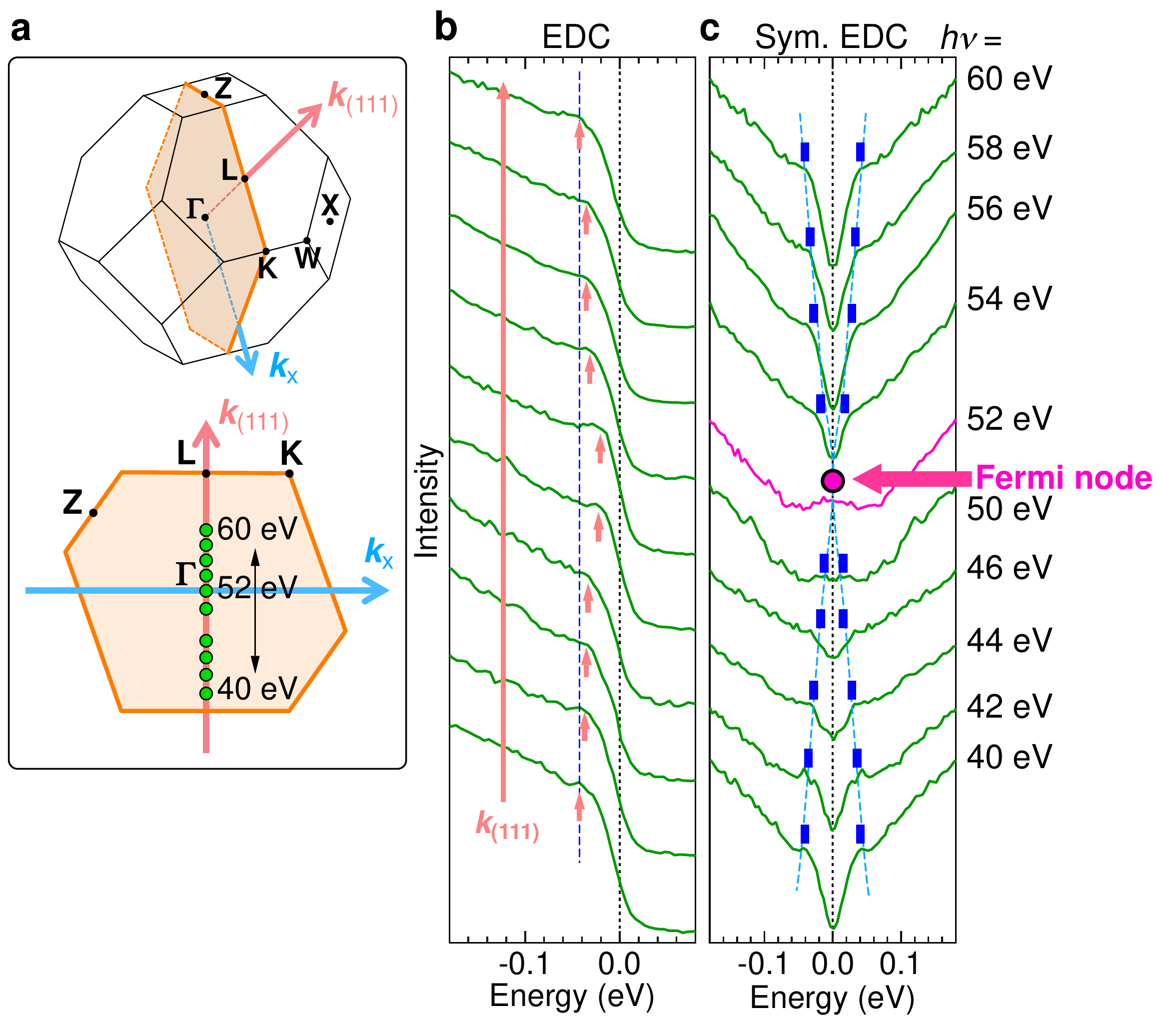}
\caption{
$\bf a,$  Brillouin zone of Pr$_2$Ir$_2$O$_7$, indicating the measured momentum plane (painted by orange) and momentum points for various photon energies (green circles). 
$\bf b,$  EDCs at various momentum points along $k_{\rm (111)}$ (green circles in $\bf a$) measured at different photon energies. The magenta arrows indicate the peak positions of the spectra. A blue dashed line is added to confirm an energy dispersion of the peaks along  $k_{\rm (111)}$. 
$\bf c,$ The symmetrized EDCs of $\bf b$. The blue bars and dashed curves indicate the spectral peaks and the energy dispersion, respectively. 
}
\label{3rd_kz}
\end{figure*}
\newpage 

\begin{figure*}  
\includegraphics[width=6in]{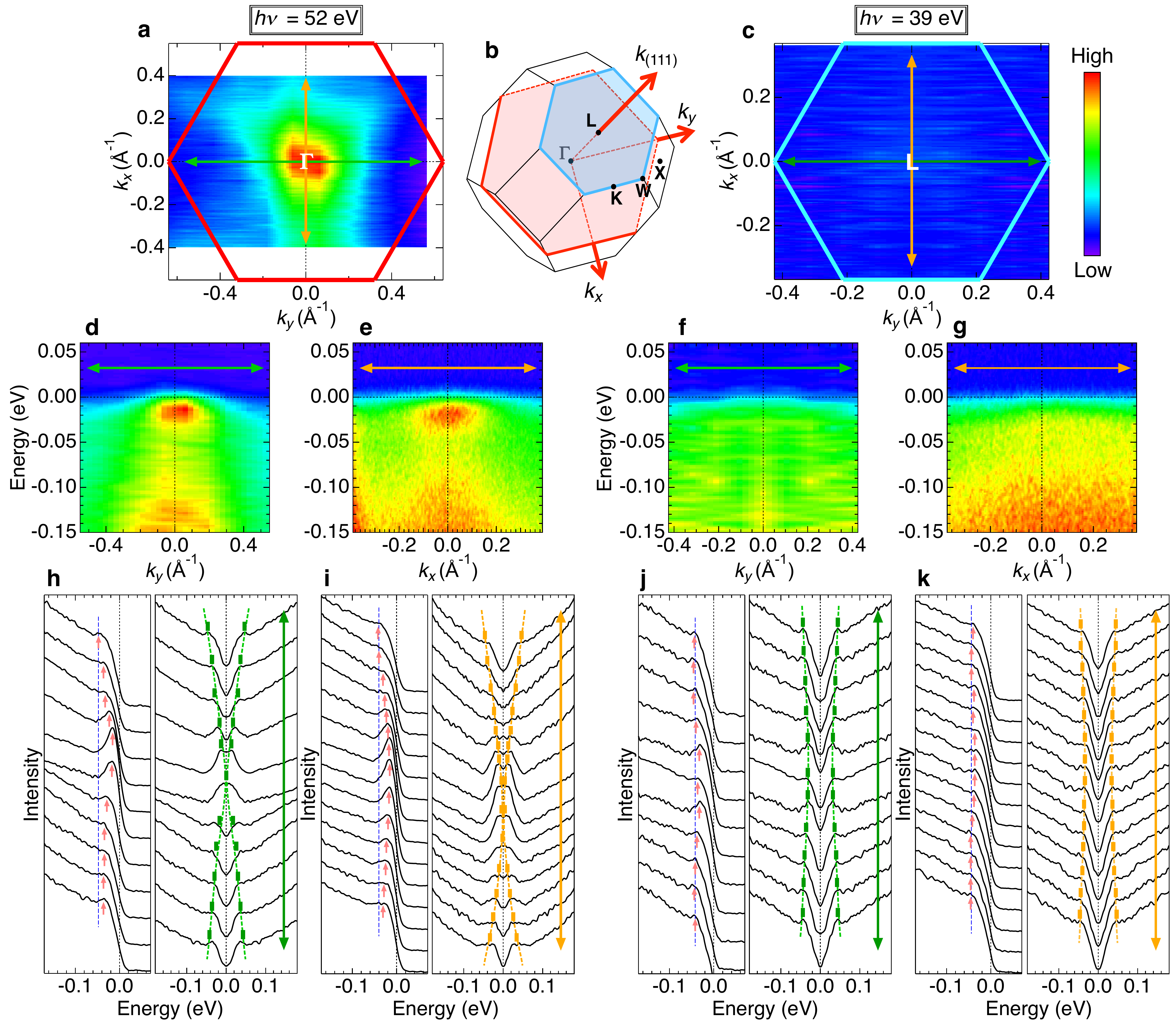}
\caption{
$\bf a,c,$ ARPES intensities at $E_{\rm F}$ along momentum sheets perpendicular to the $k_{\rm (111)}$ axis (see panel $\bf b$), crossing $\Gamma$ ($h\nu=52$ eV) and L ($h\nu=39$ eV), respectively.  For the data at $h\nu=39$ eV, only the positive side of $k_y$ were measured, and it is reflected to the negative side.
$\bf b,$ Brillouin zone, indicating the momentum sheets for the data of $\bf a$ and $\bf c$, which are painted by red and blue, respectively. $\bf d,e,$ Band dispersion maps along $k_y$ and $k_x$ marked by a green and orange arrows in $\bf a$, respectively. $\bf f,g,$ Band dispersion maps along $k_y$ and $k_x$ marked by a green and orange arrows in $\bf c$, respectively.
$\bf h,i,$ The left panels plot EDCs extracted from $\bf d$ and $\bf e$, respectively. The magenta arrows indicate  energies of spectral peaks. A blue dashed line is added to confirm the energy dispersion. To the right, the symmetrized EDCs are shown. 
Green and orange bars and dashed curves indicate the spectral peaks and the energy dispersion of those. 
$\bf j,k,$ The same data as $\bf h$ and $\bf i$, but extracted from $\bf f$ and $\bf g$, respectively. 
}
\label{3rd_BZ}
\end{figure*}
\newpage 

\begin{figure*}  
\includegraphics[width=6 in]{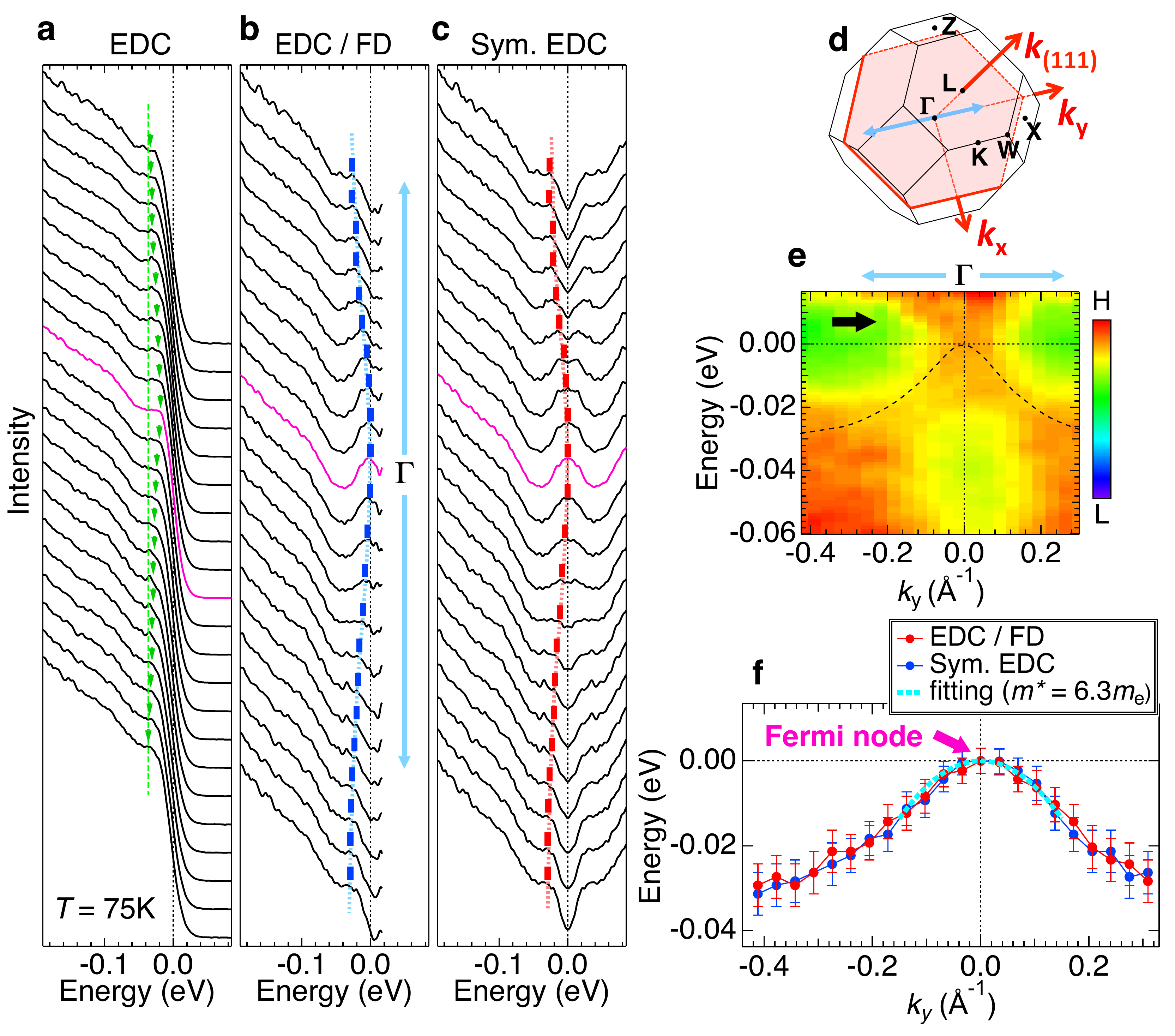}
\caption{Consistency between the results obtained with the FD-division method and the symmetrization method.
$\bf a,$ The ARPES spectra (EDCs) measured at $T=75$K along a momentum cut crossing $\Gamma$ indicated in 
$\bf d$ with a light blue arrow. The green dashed line is a guide to eyes for a confirmation of dispersion in the peaks (green arrows). 
$\bf b,$ The same EDCs in $\bf a$, but divided by the Fermi function at 75K convoluted with the experimental energy resolution. 
 $\bf c,$ The same EDCs in $\bf a$, but symmetrized about $E_F$. 
 The color bars and dashed curves in $\bf b$ and $\bf c$ indicate the spectral peaks and the energy dispersions, respectively. 
 $\bf d,$ Brillouin zone of Pr$_2$Ir$_2$O$_7$ with a measured momentum cut (a light blue arrow).  
  $\bf e,$ The corresponding ARPES image after the treatment of FD-division.  The black arrow points to the remarkable intensity above $E_F$,
  implying the existence of conduction  band touching on the unoccupied side. 
$\bf f,$ The band dispersion extracted from the peak positions of spectra in $\bf b$ and $\bf c$. 
The light blue dashed curve is a quadratic dispersion fitted to the data. Error bars in {\bf f} represent uncertainty in estimating the spectral peak positions. 
}
\label{EDCvsSym}
\end{figure*}
\newpage 

\begin{figure*}  
\includegraphics[width=6in]{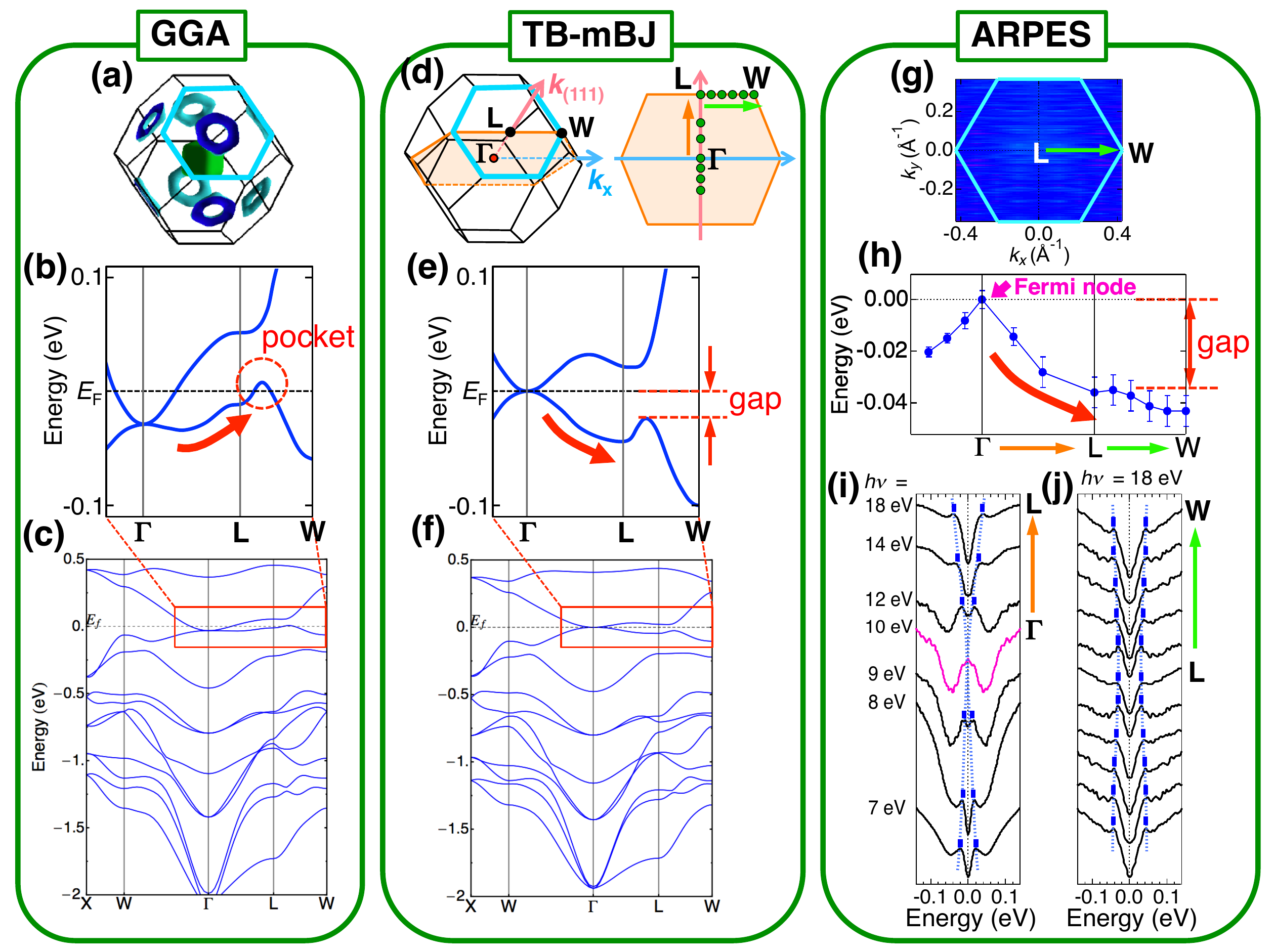}
\caption{Agreement of the ARPES data with the TB-mBJ calculation, and disagreement with the GGA calculation. 
{\bf a-c, } The results of GGA calculations: the Fermi surfaces ($\bf a$) and the energy dispersions ($\bf b,c$). 
The magnified image of red square region in $\bf c$ is exhibited in $\bf b$. 
{\bf d-f, } The results of TB-mBJ calculations: the Fermi node state ($\bf d$) and the energy dispersions ($\bf e,f$). The magnified image of red square region in $\bf f$ is exhibited in $\bf e$. 
{\bf g-j,} The ARPES results: the Fermi intensity along the $k_x-k_y$ plane crossing the L point ($\bf g$) (light blue region marked in $\bf a$ and $\bf d$), the energy dispersions determined along the high symmetry lines of $\Gamma$-L-W ({\bf h}), and the corresponding ARPES spectra (symmetrized EDCs) along $\Gamma$-L ($\bf i$) and L-W ($\bf j$).  The observed $k$ points are marked in the right panel of $\bf d$ with green circles. 
The band dispersion obtained by GGA calculation approaches $E_F$ from $\Gamma$ toward L (tick red arrow in {\bf b}), and crosses $E_F$ along the L-W cut, generating a Fermi pocket (red dashed circle in {\bf b}). 
These features disagree with the ARPES data, demonstrating that the band disperses toward higher binding energies along the $\Gamma$-L cut with a gap opening all the way toward the W point without crossing $E_F$. 
On the other hand, the TB-mBJ calculation reproduces all the features seen in the ARPES data pointing to the quadratic Fermi node state. Error bars in {\bf h} represent uncertainty in estimating the spectral peak positions.
}
\label{GGA_vs_TB-mBJ}
\end{figure*}
\newpage 

\begin{figure*}  
\includegraphics[width=6in]{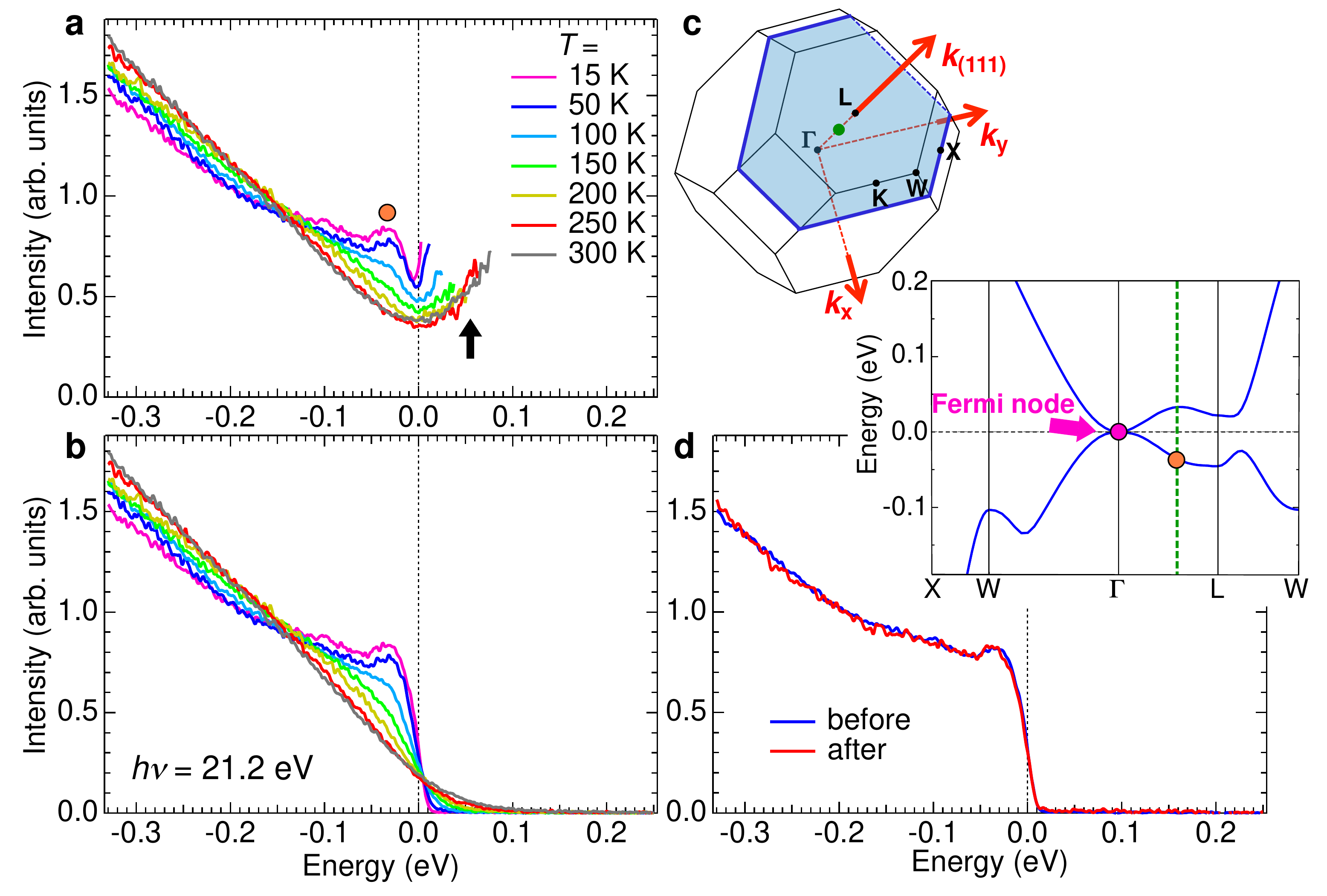}
\caption{Dramatic peak suppression at elevated temperatures, and aging check after the measurement of  temperature sweeping. $\bf a,$ Temperature evolution of EDCs divided by the energy-resolution convoluted Fermi function at measured temperatures. The 21.2 eV photons (HeI$\alpha$) were used, and the observed $k$ point is marked in $\bf c$ with a green circle, and also in the inset of $\bf d$ with a green dashed line. The peak position is marked by an orange circle. The black arrow in $\bf a$ indicates an upturn behavior of the spectra beyond $E_{\rm F}$. The data up to 150K are shown in Fig. 5 of the main paper. Here we show the data up to the higher temperature of 300K. 
$\bf b,$  EDCs used for the analysis in $\bf a$.  $\bf c,$ Brillouin zone: a momentum sheet measured at $h\nu=21.2$ eV and the $k$ point where EDCs of $\bf a$ were taken are marked by a blue sheet and a green circle, respectively. $\bf d,$  
The following two energy distribution curves are compared; one taken just after the sample cleaving and the other taken after completing the temperature scan presented in $\bf b$. These two curves almost perfectly overlap, signifying that there is no aging of sample surface during the entire measurement. 
The inset shows the calculated band dispersion along the high symmetry lines (the same figure as Fig. 5d). 
}
\label{aging}
\end{figure*}
\newpage 

\begin{figure*}  
\includegraphics[width=6in]{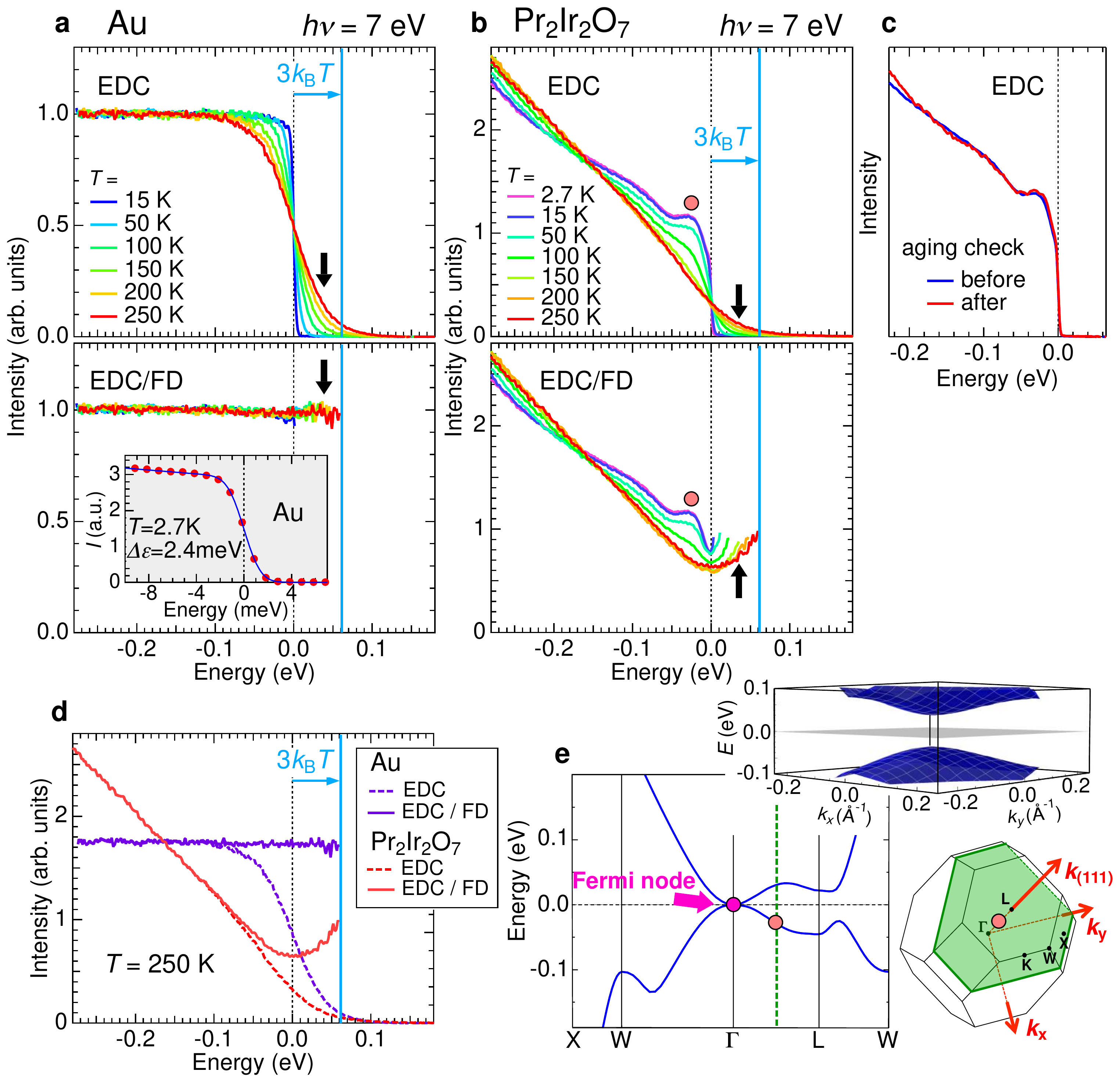}
\caption{ Signature of the unoccupied conduction band of Pr$_2$Ir$_2$O$_7$
observed by a laser ARPES with ultra-high energy resolution. 
$\bf a,$ The ARPES spectra of Au (electrically contacted to the samples) measured at various temperatures by a 7-eV laser ARPES (the upper panel). 
In the bottom panel, the same spectra, but divided by the Fermi function at the measured temperature convoluted with the experimental energy resolution, are plotted.  The energy resolution is estimated to be $\Delta \varepsilon =2.4$ meV in the present experimental setting 
as demonstrated in the inset panel using the Au spectrum at the lowest temperature ($T=2.7$K). 
$\bf b,$ Temperature evolution of ARPES spectra for the sample (the upper panel) and the same curves after the FD-division (the lower panel), measured at the midpoint between the $\Gamma$ and L points reached at $h\nu =7$eV.
The experimental setting and the FD-division method applied to the data are exactly the same as for the Au in  $\bf a$.
The arrows in $\bf a$ and $\bf b$ mark the thermally populated spectral weight above $E_F$, exhibited up to 3$k_BT$. 
$\bf c,$ The comparison of spectral shape before and after the temperature scan. $\bf d,$ Comparison between the spectra of Au and the sample (EDCs and FD-divided EDCs extracted from $\bf a$ and $\bf b$).  
$\bf e,$ Band calculation close to $E_F$. The $k_z$ (or $k_{(111)}$) value at $h\nu =7$eV is located at the midpoint between the $\Gamma$ and L points (a dashed green line and a green $k_x-k_y$ plane).
}
\label{LaserARPES}
\end{figure*}
\newpage 

\begin{figure*}
\includegraphics[width=6.3in]{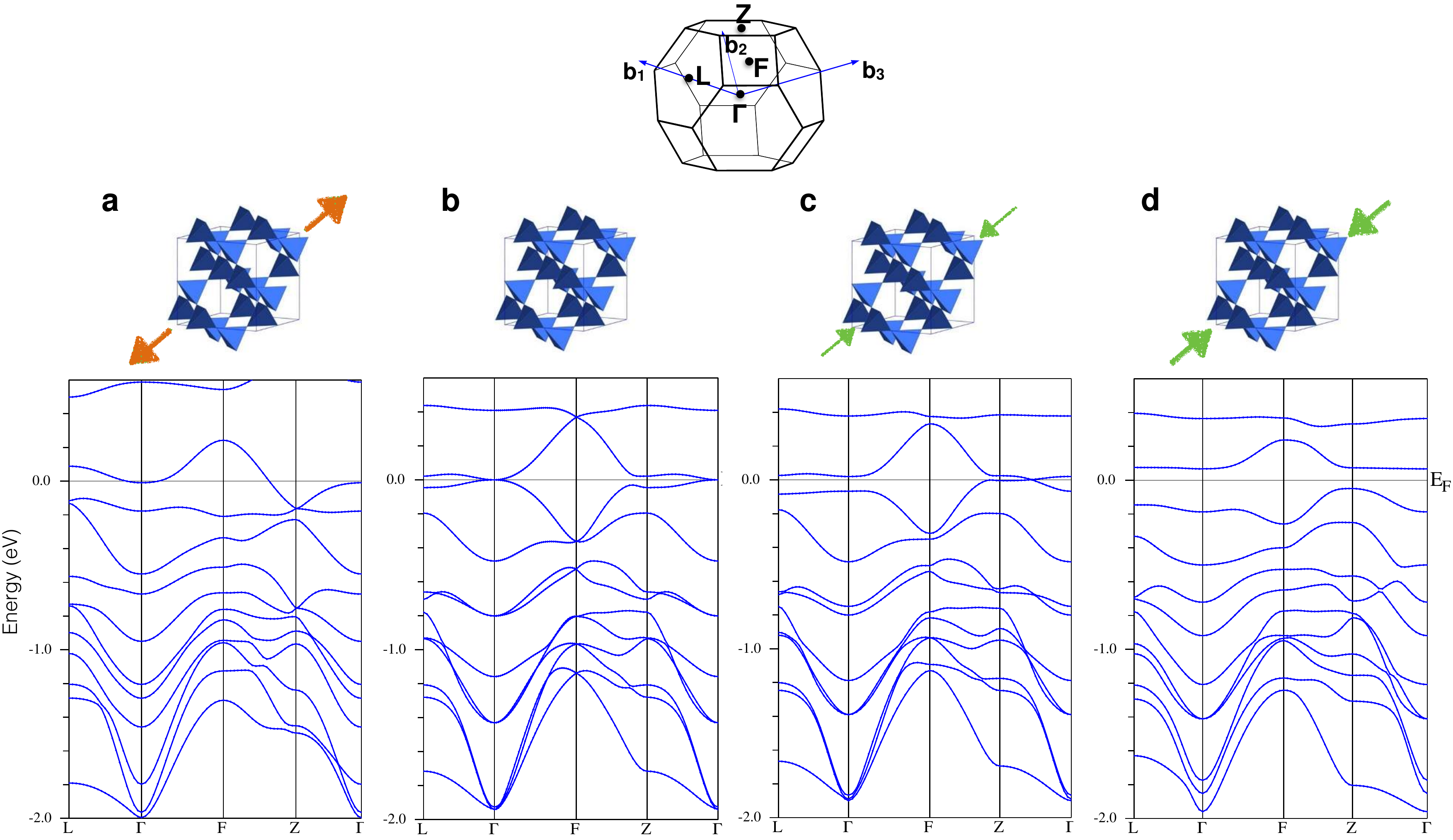}
\caption{Schematic graphs of Pr$_2$Ir$_2$O$_7$  under uniaxial pressure along $\langle 111\rangle$ direction and the corresponding Brillouin zone of the rhombohedral unit cell, with TRIM specified. {\bf a,} TB-mBJ paramagnetic band structure of Pr$_2$Ir$_2$O$_7$ under 5 $\%$ in-plane compressive strain ($a=0.95a_0$, $a_0=7.35$ \AA, the bulk in-plane lattice parameter in the hexagonal cell). The band crossing occurs along $\langle 111\rangle$ direction, which is located at the Z point ($\pi$,$\pi$,$\pi$). {\bf b,} TB-mBJ paramagnetic band structure of Pr$_2$Ir$_2$O$_7$ bulk in the absence of pressure or strain in the rhombohedral unit cell. {\bf c,} TB-mBJ paramagnetic band structure of Pr$_2$Ir$_2$O$_7$ under 1 $\%$ in-plane tensile strain ($a=1.01a_0$). A tiny gap, 4 meV is opened, which can not be oberved from the scale shown. {\bf d,} TB-mBJ paramagnetic band structure of Pr$_2$Ir$_2$O$_7$ under 5 $\%$ in-plane tensile strain ($a=1.05a_0$).  The gap is clearly shown. } 
\label{supple-fig2}
\end{figure*}

\begin{figure*}
\vspace {1.5cm} 
\includegraphics[width=5.5in]{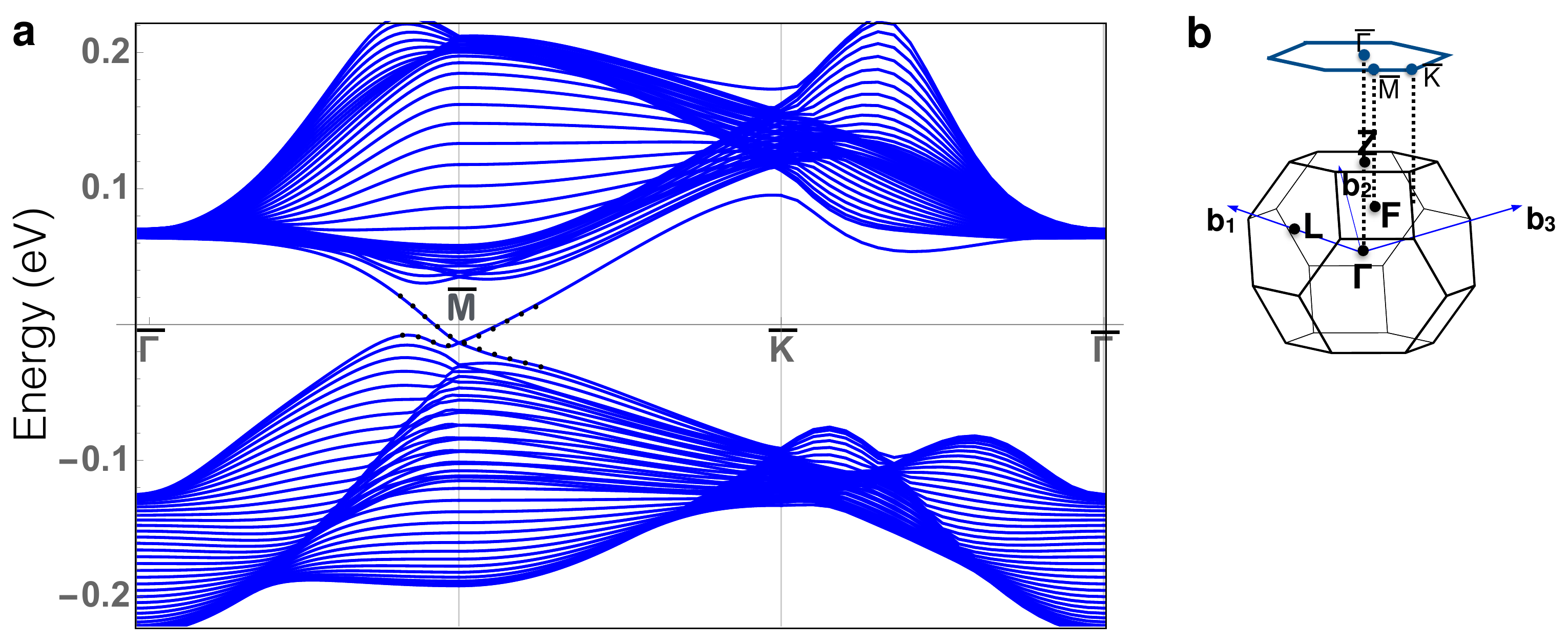}
\caption{{\bf a,} The paramagnetic band structure of Pr$_2$Ir$_2$O$_7$ under 5\% uniaxial strain on the $\langle 111\rangle$ surface. The topological surface state is marked with dotted lines and the Dirac cones are located at the $\overline{\textrm M}$ point. {\bf b,} The surface 2D Brillouin zone is shown by the blue hexagon, in which the high symmetric $k$ points $\overline{\Gamma}$, $\overline{\textrm M}$ and $\overline{\textrm K}$ are labeled. } 
\label{supple-fig4}
\end{figure*}
